\documentclass[usenames,dvipsnames, 10pt,preprint2]{emulateapj}
\usepackage{amsmath}
\slugcomment{Draft Version \today}

\usepackage{times}
\usepackage{graphicx}
\usepackage{verbatim}
\usepackage{tikz}
\usepackage{color}

\usepackage[T1]{fontenc}
\usepackage{aecompl}
\usepackage{calligra}
\DeclareMathAlphabet{\mathcalligra}{T1}{calligra}{m}{n}
\DeclareFontShape{T1}{calligra}{m}{n}{<->s*[2.2]callig15}{}

\usepackage{hyperref}
\hypersetup{
    pdfnewwindow=true,      
    colorlinks=true,       
    linkcolor=blue,          
    citecolor=blue,        
    filecolor=blue,      
    urlcolor=blue           
}

\def\prd{PRD}
\def\apj{ApJ}                 
\def\apjl{ApJL}               
\def\apjs{ApJS}               
\def\mnras{MNRAS}             
\def\aap{A\&A}                
\def\aj{AJ}                   
\def\nat{Nature}              
\def\araa{ARA\&A}             
\def\ssr{SSR}                 
\def\nar{NewAR} 			  

\def\bin{\rm{bin}}
\def\var{\rm{var}}

\def\obs{\rm{obs}}
\def\res{\rm{res}}

\def\min{\rm{min}}
\def\max{\rm{max}}
\def\Edd{\rm{Edd}}
\def\GW{\rm{GW}}

\def\EHT{\rm{VLBI}}
\def\Msun{{M_{\odot}}}

\def\mde{\dot{\mathcal{M}}}
\def\base{\rm{base}}
\def\trans{\rm{trans}}
\def\GW{\rm{GW}}

\def\Pmin{\rm{Pmin}}
\def\Vmin{\rm{Vmin}}

\def\be{\begin{equation}}
\def\ee{\end{equation}}

\def\bea{\begin{eqnarray}}
\def\eea{\end{eqnarray}}

\begin{document}
\title[Imaging MBHB Orbits]{Repeated Imaging of Massive Black Hole Binary Orbits with Millimeter Interferometry: \\
measuring black hole masses and the Hubble constant}
\author{Daniel J. D'Orazio}
\email{daniel.dorazio@cfa.harvard.edu}
\author{Abraham Loeb}
\affiliation{Department of Astronomy, Harvard University, 60 Garden Street Cambridge, MA 01238, USA}

\begin{abstract} 
Very long baseline interferometry (VLBI) at millimeter (mm) wavelengths is
being employed to resolve event-horizon scale structure of the
environment surrounding the Milky-Way black hole, at an angular resolution of
a few tens of micro-arcseconds. The same approach could also resolve the
orbital separation of a population of massive black hole binaries (MBHBs).
Modeling the inspiral of binaries due to gravitational wave emission and gas
and requiring binary orbital periods of less than 10 years, we estimate that
there may exist $\sim100$ resolvable MBHBs that are bright enough to be
observed by mm-wavelength VLBI instruments over the entire sky, at redshifts
$z\lesssim0.5$. We propose to search for these resolvable MBHBs by identifying
binaries with the required orbital separations from periodic quasar light
curves identified in optical and near-IR surveys. These  periodic-light-curve
candidates can be followed up with radio observations to  determine their
promise for observation with VLBI at mm wavelengths. VLBI observations over
the timescale of a binary orbit can allow unprecedented precision in the
measurement of the binary mass, to within $30\%$. In combination with an
independent binary mass measurement, VLBI observation would allow a novel
$\mathcal{O}(10\%)$ measurement of the Hubble constant, independent from those
currently proposed and employed.
\end{abstract}

\maketitle

\section{Introduction}

The merger of galaxies harboring massive black holes \citep[MBHs][]{kr95,
KH2000, ff05, KormendyHo2013} can lead to the formation of a compact massive
black hole binary (MBHB) at the center of the newly formed galaxy
\citep{ColpiDotti:2011:rev}. For moderate MBH mass ratios
($\gtrsim$ 1:100), dynamical friction can bring the MBHs together on the
galactic dynamical timescale to form a hard binary with orbital separation of
order parsecs (pcs) \citep[\textit{e.g.}][]{Callegari:2011,
Mayer:2013:MBHBGasRev, DosopoulouAntonini:2016}. Dynamical friction becomes
inefficient at hardening the binary further at smaller orbital separations and
alternative mechanisms for removing binary angular momentum and energy must be
employed if the binary is to shrink to $\lesssim 0.01$ pc separations, where
gravitational radiation will bring the binary to coalescence
\citep[\textit{e.g}, ][]{Begel:Blan:Rees:1980, MerrittMilos:2005:LRR}.

Multiple mechanisms are capable of shrinking the binary orbital separation
through this intervening stage, solving the  so-called final-parsec problem
\citep{Milosavljevic:2003:FPcP}. Possible solutions include interaction of the
binary with a gas disk \citep{GouldRix:2000, ArmNat:2002:ApJL} \citep[for
recent work see][and references therin]{Yike+2017}, a massive perturber
\citep{Goicovic+2016}, or non-axisymmetric stellar distributions that allow a
high interaction rate between stars and the binary \citep[see][and references
therein]{Gualandris+2017:TriaxRefill}. However, to truly understand the
mechanisms that drive MBHBs to merge in galactic nuclei, we must
find observational tracers of the MBHB population, probing different stages of
MBHB evolution. Based on reliable tracers of the MBHB population, the relative
fraction of MBHBs at different orbital separations can be translated into the
rates at which the binaries are driven together during these stages, and hence
elucidate the mechanisms driving orbital decay \citep[see,
\textit{e.g.},][]{HKM09}.

These observational tracers naturally fall into the realm of multi-messenger
astronomy. The earliest stages of MBHB formation, where the MBHs have not yet
hardened into a binary, have been captured by direct electromagnetic (EM)
imaging; two distinct active galactic nuclei with projected
separations of 10's to 1000's of pcs have been identified in radio, optical,
and X-ray wavelengths \citep{Komossa:Rev06, Rodriguez:2006, BurkeSpolaor:2011,
Fabbiano:2011, Dotti:2012:rev, Civano+2012, Blecha+2013, Comerford:2013,
Woo:subKpcBin:2014}. Low frequency gravitational radiation will probe the
final stages of a MBHB's life. Particularly, the stochastic background of
gravitational waves detectable by the Pulsar Timing Arrays (PTAs)
\citep[][]{PTAs, ManchesterIPTA:2013, NANOGrav:2013, EPTA:2013, PPTA:2013,
Shannon:2015}, will be sensitive to environmental effects determining the
binary eccentricity and lifetime at $\sim$nHz orbital frequencies
\citep{Kelley+2017}, probing the late inspiral of the most massive MBHBs.

The early, dynamical-friction-driven, and late, gravitational-radiation-driven
phases of MBHB evolution are separated by the sub-pc orbital separation
regime. At  sub-pc orbital separations, it is likely that gas will accompany
the MBHB \citep[\textit{e.g.}][]{Barnes:1996, Barnes:2002}, not only aiding in
resolution of the final-pc problem, but also providing the potential for
bright EM signatures. However, identification of an EM signature with a
compact MBHB must surmount the obstacle of disentangling signatures of an
accreting MBHB from those of a single, accreting MBH. Finding unique EM
identifiers of accreting, compact MBHBs has been the subject of numerous
theoretical studies and corresponding observational searches. Possible
signatures could arise from emission line dynamics
\citep[\textit{e.g.},][]{ShenLoeb:2010, Tsalmantza:2011, Bogdanovic+2009,
Eracleous+2012, McKFeZoltan:2013, DecarliDott:2013:SpecMBHBcandI, Shen+2013I,
Liu+2014II, LiuEracHalp:2016}, tidal disruptions by a binary
\citep[\textit{e.g.},][]{LiuChen:2009, StoneLoeb:2011, Coughlin+2017},
peculiar jet morphology \citep[\textit{e.g.},][]{Gower+1982, Roos:1993,
MerrittEker:2002, Zier:2002, Romero:2000, Kun+2014, Kun+2015:PG1302,
KulkLoeb:2016}, orbital motion of an unresolved radio core observed with VLBI
\citep[][similar in goal to the ideas discussed here]{Sudou+2003},
or periodic emission and lensing events from quasars \citep{Hayasaki:2008,
HKM09, DHM:2013:MNRAS, PG1302MNRAS:2015a, Farris:2014, DDLens:2017}.
Observational searches motivated by these studies have identified a number of
individual MBHB candidates \citep{Bogdanovic+2009:MBHBcand, Valtonen+2008,
LiuKomossa:2014, Graham+2015a, Li+2017:ARk120}, and recently, time domain
searches for periodically varying quasars have identified $\sim 140$ MBHB
candidates \citep{Graham+2015b, Charisi+2016}.

In order to use the growing population of MBHB candidates to investigate the
drivers of MBHB evolution, further vetting of these candidates must be
employed. For this to happen, new ways of identifying compact MBHBs, in
conjunction with existing methods, must be developed. In this work we suggest
the use of very long baseline interferometry (VLBI) at millimeter (mm)
wavelengths to directly image the sky-projected orbital path of sub-pc
separation MBHBs.

The outline of this paper is as follows. In Section \ref{S:Imaging} we present
our criteria for tracking a MBHB orbit with VLBI. In Section \ref{S:MBHBPop}
we present our calculation of the population of resolvable MBHBs and the
resulting gravitational wave background due to this population.
We present results in Section \ref{S:Results}. In
Section \ref{S:Results:a} we present the our main results while
Sections \ref{S:Results:b} and \ref{S:Pdeps} provide a detailed analysis of
the dependence of our results on model parameters (the reader primarily
interested in the main results and implications may wish to skip Sections
\ref{S:Results:b} and \ref{S:Pdeps}). In Section \ref{S:Discussion} we
discuss the application of MBHB imaging to inferring the MBHB population,
measuring the binary mass, and measuring the Hubble constant. In Section
\ref{S:Conclusions} we conclude.

\section{Imaging the orbit of a compact MBHB}
\label{S:Imaging}

To image a MBHB orbit, we require that the binary orbital separation be i)
larger than the minimum spatial resolution, ii) larger than the size of the
emission region at the observing wavelength, and iii) both binary components
be bright enough to be detectable independently, or that
one component be bright and a calibrator source be nearby. We additionally
impose that the binary orbital period be shorter than some maximum baseline
timescale, $P_{\base}$. By observing an entire orbit, we ensure that the
binary nature of the source can be determined.

The first criterion can be met by mm-wavelength VLBI. VLBI experiments with
maximum baselines the size of the Earth can reach diffraction limited
resolutions on the order of $20\mu$as when observing in $\lesssim$mm
wavelengths, and can reach sub-diffraction limited, down to $4\mu$as,
resolutions using novel image reconstruction techniques \citep{Akiyama+2017a,
Akiyama+2017b}.\footnote{This fact is a leading driver behind the Event
Horizon Telescope, which is currently being employed to resolve 
Schwarzschild-radius scale structure of the environment surrounding the Milky-Way 
black hole \citep{Doeleman+2008}.} Astrometric tracking of a source can reach 1$\mu$as
precision \citep{BrodLoebReid:2011A}. At 1 Gpc, a $10\mu$as resolution
corresponds to a physical binary orbital separation of $0.05$pc and an orbital
period of only $10$ years for the most massive, $10^{10} \Msun$, binaries.
Hence the first criteria can be satisfied because radio-loud active galactic
nuclei (AGN), which may harbor close MBHBs, are also bright at mm and sub-mm
wavelengths \citep[\textit{e.g.}][]{Elvis+1994}.

To determine the validity of the second criterion, we estimate the size of the
mm-wavelength emission region. Binaries for which this region is smaller than
the separation will be viable targets for sub-mm VLBI imaging, otherwise the
photosphere of the mm-emission region could mask the resolvable binary
components or emanate from a region in a jet that is larger than the orbital
separation. While emission regions that are larger than the binary orbital
separation may still provide evidence for a binary via photometric variability
or periodically changing geometry, they are possibly more
complicated than the case of two distinguishable sources envisioned here.

Observationally, we can probe the size of the mm- to sub-mm emission region
from variability measurements. Specifically, if the emission region is
smaller than the binary separation, then in the most conservative case,
causality requires that the light-travel distance over the duration of the
shortest mm-variability timescales be smaller than the binary orbital
separation,
\begin{equation}
\frac{c \Delta t^{\rm{min}}_{\var}}{(1+z)} \leq  \theta_{\min} D_A(z) \leq \left(\frac{ \sqrt{GM} P_{\base}}{2 \pi (1+z)} \right)^{2/3} ,
\label{Eq:VarTimes}
\end{equation}
where the middle term is the smallest possible binary separation and the
rightmost term is the largest binary separation for maximum allowed binary
orbital period $P_{\base}$. $D_A(z)$ is the angular diameter distance of the
MBHB at redshift $z$.

Equation (\ref{Eq:VarTimes}) requires that the observed mm-variability timescales
satisfy $\Delta t^{\rm{min}}_{\var} \lesssim  1 \ \rm{day} \
(\theta_{\min}/1 \mu\rm{as})$ to resolve all possible MBHBs at $z\geq0.02$ or
$\Delta t^{\rm{min}}_{\var} \lesssim  54 \ \rm{days} \ M^{1/3}_9 P^{2/3}_{10}$
to resolve only the longest period, and most common (see next section) MBHBs
at $z>0.02$. Here $M_9$ is the total binary mass in units of $10^9 \Msun$ and
$P_{10}$ is the maximum baseline period in units of 10 years.

Recent studies have employed the SMA calibrator database to characterize AGN
variability in the sub-mm regime \citep{Strom+2010, BowerDexter+2015}. They
quantified the variability timescale by the damped random walk correlation
timescale \citep{Macleod+2010}, finding that sub-mm variability of these
brightest sources has characteristic timescales of $\sim 1-1000$ days.
Notably, \cite{BowerDexter+2015} find that the low-luminosity AGN (LLAGN)
exhibit shorter timescale variability than other blazars and AGN in the SMA
calibrator sample. Furthermore, the characteristic timescale for variability
of these sources appears to track a multiple of the MBH innermost stable
circular orbit (ISCO), suggesting that mm-emission from LLAGN tracks the
regions very close to the MBH.

That the mm-emission from LLAGN tracks event horizon scales is consistent with
standard models for synchrotron emission from jets \citep{BK79}.
In these models, the BH launches a jet and the mm-emission
is generated by synchrotron radiation at shocks along the length of the jet.
There is a smallest distance along the length of the jet from which optically
thin, bright synchrotron radiation can be emitted. This minimum size scales
with the bolometric jet luminosity. Because the jet is launched from a small
region that is bound to the BH, the mm-emission will necessarily track the BH
orbit, regardless of its size compared to the Roche radius. Hence,
the size of the mm-emission region need not be truncated close to the BH for
VLBI-orbit tracking to be viable. Rather, because we wish to consider systems
for which the mm-emission regions emanating from each BH are clearly
distinguishable, we compare the size of the emission region with the binary
separation. We compute the size of the  mm-emission region as a function of
AGN luminosity and Eddington ratio (See Appendix \ref{S:mmSize}) and so
determine which MBHBs have mm-emission regions larger than their binary
separation. We exclude these from the population estimates below.

The final criterion, that both binary components be bright, is not only a
sensitivity issue (which we address in the next section) but a matter of
calibration necessary for VLBI. If only one binary component is bright enough
to detect, its orbital path cannot be tracked without a bright source
(within $\sim1^{\circ}$) for phase reference
\citep[\textit{e.g.}][]{BrodLoebReid:2011A}, \textit{i.e.},
the required $\sim\mu$as astrometric precision is only possible via relative
astrometry. We can make a crude estimate for the probability of finding a
bright source within $1^{\circ}$ of the target source from the number of ALMA
calibrator sources. Taking that there are about 2000 adequate calibrator
sources that could be used as phase references, we can estimate a lower limit
on the alignment probability by assuming that these calibrators are
distributed isotropically on the sky. Then the probability of finding a
suitable phase reference within $1^{\circ}$ of the source is $2000/(41252
\mbox{deg}^2) \approx 0.05$. This is non-zero, but not large enough to be
reliable. If instead the number of calibrators can be increased by a factor of
10, the probability of finding a nearby phase reference is considerable,
$50\%$. 

In the case that both binary components are bright in mm-wavelengths, the
problem is eliminated as each component can phase reference its companion.
Because we do not know the fraction of binaries for which both components are
mm-bright, and how this depends on binary parameters, AGN type, or other
unknowns, we parameterize this uncertainty with $f_{**}$.

\section{A Population of Resolvable MBHBs}
\label{S:MBHBPop}

\subsection{Calculation}
\label{S:Calc}

We next estimate the number of MBHBs that are emitting bright, mm-wavelength
radiation due to accretion, and that have an orbital separation large enough
to be resolvable by an Earth-sized VLBI array, but small enough to have a
period observable in a human lifetime.

We assume that a fraction $f_{\bin}$ of mm-bright active galactic nuclei (AGN)
are synonymous with accreting MBHBs. While this fraction is not robustly
constrained, a number of theoretical arguments imply that its value may be of
order unity \citep{KH2000, Hopkins2007a}. Additionally, the quasar lifetime
\citep{PMartini:2004} is in agreement with the time for a binary to migrate
from the edge of a gravitationally stable gas disk down to merger via gas
torques and gravitational wave losses \citep[][however, the LLAGN lifetime 
may be $\sim10-100 \times$ longer \citep{HopkinsLidz+2007}]{HKM09}. 
Also, recent searches for MBHB candidates
as periodically variable quasars estimate values of $f_{\rm{bin}} \sim 0.3$
\citep{PG1302MNRAS:2015a, Charisi+2016} from the fraction of candidates found
at a given binary period. We compute our own constraints on the binary
fraction, of the population of low-luminosity AGN considered here, in \S
\ref{S:GWB} below.

We calculate the time that a MBHB with total mass $M$ spends at a given
orbital period $P$ during the bright AGN phase. We assume that gas and
gravitational radiation drive the binary to merger to compute a residence time
at binary separation $a$,
\begin{eqnarray}
    t_{\rm{res}} &\equiv& \frac{a}{\dot{a}} = 
    \begin{cases} 
     \frac{20}{256}  \left(\frac{P}{2 \pi}\right)^{8/3} \left(\frac{GM}{c^3}\right)^{-5/3} q^{-1}_s & P < P_{\rm{trans}}  \\
      \frac{q_s}{4 \mde} t_{\Edd} & P \geq P_{\rm{trans}} \nonumber
     \end{cases}  \\
     P_{\rm{trans}} &=& 2 \pi \left( \frac{16}{5} \right)^{3/8} \left(\frac{G}{c^3}\right)^{5/8} q^{3/4}_s M^{5/8}  \left(\frac{\mde}{t_{\Edd}} \right)^{-3/8} ,
     \label{Eq:FGWGas}
\end{eqnarray}
where the first term is the residence time due purely to gravitational wave
decay \citep{Peters64}, and the second term is a prescription for orbital
decay due to gaseous effects given by \cite{Loeb:2010}. Here $q_s \equiv
4q/(1+q)^2$ is the symmetric binary mass ratio, where the standard mass ratio
is given by $q\equiv M_2/M_1$; $M_2\leq M_1$; $M_2+M_1=M$. The Eddington time,
$t_{\Edd} \equiv M/\dot{M}_{\Edd} \sim 4.5 \times 10^{7}$ yr, is the time it
takes to accrete a binary mass of material $M$ at the Eddington accretion
rate, $\dot{M}_{\Edd}\equiv L_{\Edd}/(\eta c^2)$, assuming an accretion efficiency
of $\eta=0.1$.

In the gas-driven case, the simple assumption is that the
binary orbit shrinks via interaction with the environment, either by gas
accretion, or application of positive torque to a circumbinary disk
\citep[\textit{e.g.},][]{Rafikov:2016}. Because this
rate is uncertain \citep[even its sign, \textit{e.g.},][]{Yike+2017,
MirandaLai+2017}, we parameterize the gas-driven orbital decay rate in terms
of an Eddington rate $\mde = \dot{M}/\dot{M}_{\Edd}$. The parameter $\mde$
controls the rate at which the binary orbit decays ($\dot{a} \propto \mde$ in
Eq.~\ref{Eq:FGWGas}), not the accretion rate that determines the accretion
luminosity. Hence even for the case of LLAGN, which may not experience gas
inflow at the Eddington rate, we still consider mechanisms that drive the MBHB
together at a rate comparable to if the binary torques were expelling gas at
the Eddington rate. Essentially, due to uncertainties in binary orbital decay
rates, we have purposefully not locked together the accretion mechanism and
the binary decay mechanism, we have simply parameterized the decay rate in
terms of an Eddington rate. The transition orbital period $P_{\rm{trans}}$
delineates gas-driven and gravitational-wave-driven orbital decay.

In the gas-driven case, the simple assumption is that the binary shrinks by
applying positive torque to a circumbinary disk. Then the fraction of the
inflowing gas that is expelled by the binary, and hence the binary decay rate,
is parameterized in terms of the Eddington rate at which gas can be supplied
to the binary $\mde = \dot{M}/\dot{M}_{\Edd}$. The parameter $\mde$ represents
the rate at which gas is expelled by binary torques, not the accretion rate
that determines the accretion luminosity. The transition orbital period
$P_{\rm{trans}}$ delineates gas-driven and gravitational-wave-driven orbital
decay.

From the binary residence time, we generate a probability distribution
function $\mathcal{F}(M,z)$ that provides the probability that a quasar at
a given redshift $z$, and luminosity $L = f_{\Edd}L_{\Edd}(M)$ harbors a MBHB with 
orbital period in the specified VLBI range. This probability function is derived by
integrating the residence time in Eq. (\ref{Eq:FGWGas}) over periods and mass
ratios which meet the minimum VLBI separation requirement and the maximum
period requirement, $P_{\rm{base}}$, and normalizing by the same integral over
all possible binary parameters,
\begin{eqnarray}
\mathcal{F}(M,z) &=& f_{\bin} \frac{ \int^{1}_{q^{\Vmin}_s} \int^{P_{\rm{hi}}}_{P_{\rm{lo}}} t_{\res}(M, q_s, P) \ dP \ dq_s }{  \int^{1}_{q^{\Pmin}_s} \int^{P_{\max}}_{0} t_{\res}(M, q_s, P) \ dP \ dq_s  }\nonumber \\
P_{\rm{lo}} &=&  \frac{2 \pi \left( \theta_{\min} D_A(z) \right)^{3/2}}{\sqrt{GM}} \nonumber \\
P_{\rm{max}} &=& \frac{2 \pi a^{3/2}_{\max} }{\sqrt{GM}} \nonumber \\
P_{\rm{hi}} &=& \min( P_{\base}, (1+z)P_{\max} ).
\label{Eq:BHBProb}
\end{eqnarray}
The normalization introduces three additional parameters. The first two are
the minimum (symmetric) mass ratio of the entire MBHB population
$q^{\Pmin}_s$, and the minimum for the resolvable population $q^{\Vmin}_s$.
We adopt a flat distribution in mass ratio and fiducially
set the two equal to $0.01$, a value motivated by the minimum mass ratio for
which dynamical friction can form a central binary \citep{Callegari:2011,
Mayer:2013:MBHBGasRev}. We also vary $q^{\Vmin}_s$ to larger values to
determine the mass ratio dependence of our results. We note
that $q^{\Vmin}_s$ could have a dependence on binary mass, for example
observationally through the Eddington luminosity and flux sensitivity.
However, in accordance with our choice of a flat mass ratio distribution, we
do not explore this possibility here.

The third new parameter is $a_{\max}$, the maximum binary separation for which
radio-loud quasar activity is triggered. This is required because the
residence time due to gas accretion (large $a_{\max}$) is independent of the
binary separation (the binary spends equal time per $\ln{a}$ in the gas-driven
phase), and hence, we cannot simply set an $a_{\max}$ in the normalization to
corresponds to a quasar (or LLAGN) lifetime. As noted above however, the
observationally inferred AGN lifetime is similar to that required for a MBHB
to migrate through a gas disk with an outer edge set by the Toomre stability
limit \citep{Goodman:2003, HKM09}, where the gas disk fragments into stars. We
use this separation, corresponding to the outer edge of a gravitationally
stable disk, to motivate fiducial parameter choices below.

The number of MBHBs out to redshift $z$, over the entire sky,
with binary separation resolvable by a VLBI array, and with
orbital period limited by $P_{\base}$ is,
\begin{eqnarray}
N_{\EHT}
&\approx& 4 \pi \int^{z}_0  \frac{d^2V}{dz d\Omega} \int^{\infty}_{L^{\min}_{mm}}{\frac{d^2N}{dL_{mm} dV} \mathcal{F}(\boldsymbol{\chi}; L_{mm}, z) dL_{mm}} dz \nonumber \\
&\boldsymbol{\chi}& = (\theta_{\min}, a_{\max}, \mde, f_{\Edd}, P_{\base}, q^{\Vmin}_s, f_{\bin}, f_{**}) ,
\label{Eq:NEHT}
\end{eqnarray}
where, from left to right, we incorporate the cosmological volume element in a flat
universe \citep[\textit{e.g.}][]{HoggCosmoDist:1999}, a mm-wavelength AGN
luminosity function (mmALF; see Appendix \ref{S:mmALF}), and the binary probability
distribution function discussed above. We have re-written the binary
probability in terms of mm-wavelength luminosity through the relation,
\begin{equation}
M = \frac{ L_{\rm{bol}}(L^{\obs}_{mm})\sigma_T}{f_{\Edd} 4 \pi G m_p c } \Msun ,
\label{Eq:MofL}
\end{equation}
where we have assumed that the accretion on to the binary generates bolometric
luminosity equal to a fraction $f_{\Edd}$ of the Eddington Luminosity
($L_{\Edd} = 4 \pi G M m_p c /\sigma_T$) and we estimate the bolometric
luminosity from the observed mm-wavelength luminosity (see below).

\begin{figure}
\begin{center}
\includegraphics[scale=0.22]{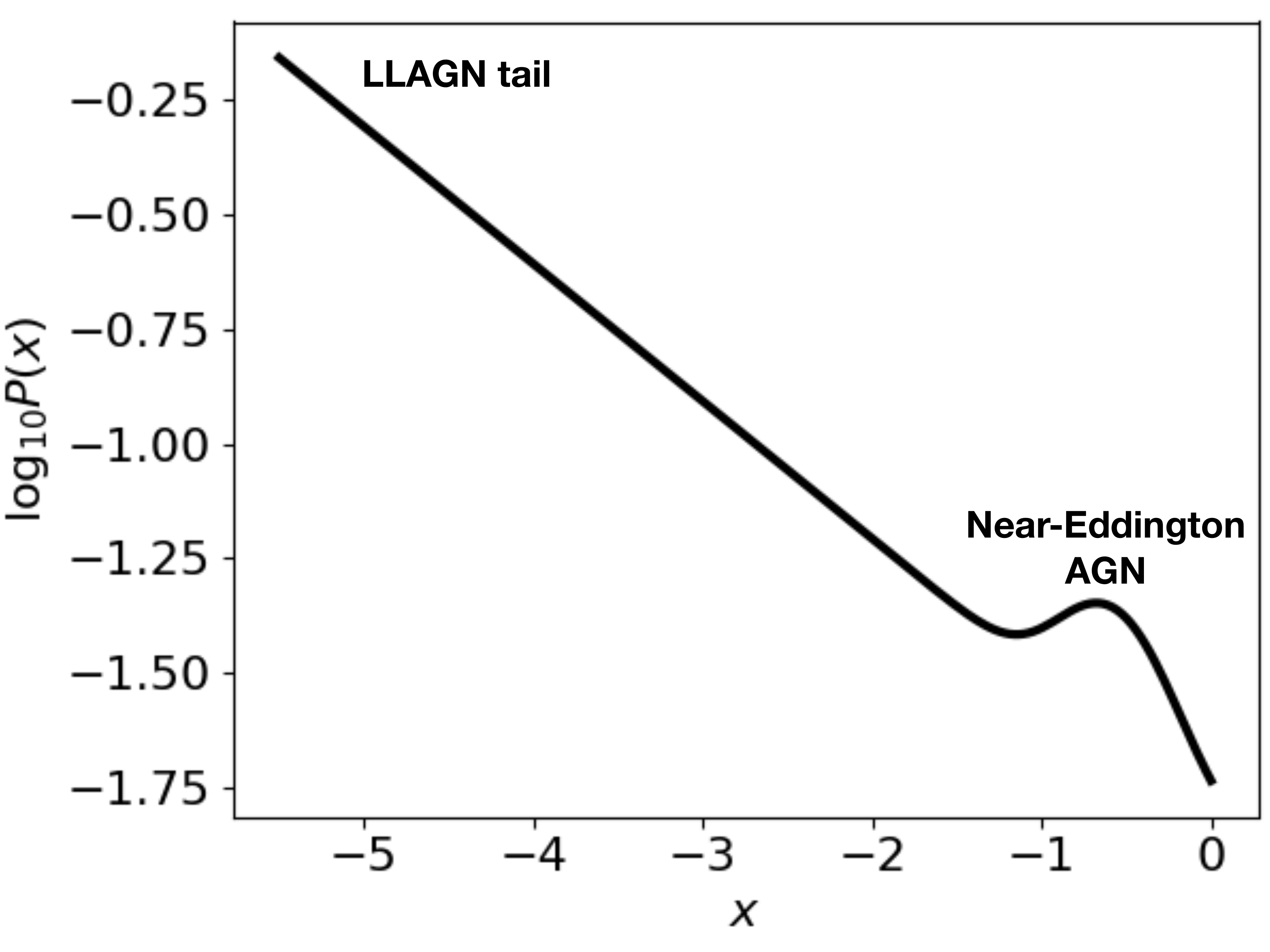}
\end{center}
\vspace{-20pt}
\caption{
The distribution, Eq. (\ref{Eq:fEddDist}), plotted in terms of $x \equiv \log_{10}
f_{\Edd}$.
}
\label{Fig:Pofx}
\end{figure}

In choosing $f_{\Edd}$, recall that LLAGN are expected to have
mm-wavelength emission emanating from a small region around each BH, making
them  well-suited for the imaging study proposed here. To incorporate LLAGN we
choose a distribution $P(x)$ of the log Eddington fraction, $x \equiv \log_{10}
f_{\Edd} \leq 0$, that consists of a power law with a slope $a=-0.3$ and
minimum value of $x_{\min} = -5.5$, plus a Gaussian in $x$ with mean at $x_0 =
-0.6$ and standard deviation $\sigma = 0.3$,
\begin{equation}
P(x) = \frac{ \left(10^x \right)^a + \frac{1}{\sqrt{2 \pi } \sigma} \exp{\left[-(x-x_0)^2/(2 \sigma^2)\right]}}{\frac{1}{2} \left[ \rm{erf}\left(  \frac{x_0-x_{\min}}{\sqrt{2}\sigma} \right) - \rm{erf}\left(   \frac{x_0}{\sqrt{2}\Sigma} \right) \right] + \frac{1 - 10^{a x_{\min}}}{a \ln{10}} }.
\label{Eq:fEddDist}
\end{equation}
We plot $P(x)$ in Figure \ref{Fig:Pofx}. This choice is based on
observations of a normally distributed population of AGN accreting near
Eddington, and a power law tail of LLAGN \citep[][but see also
\cite{Weigel+2017_fEdds}]{KauffHeck:2009, SWM:2013_fEdds}. The value of
$x_{\min}$ is based on Figure 5 of \cite{Eracleous:LLAGN+2010}. While this
$f_{\Edd}$ distribution comprises our fiducial model, we also employ a simpler
delta function $f_{\Edd}$ distribution for comparison.

The lower limit on the observed, specific luminosity is written in terms of
the specific flux sensitivity $F^{\min}_{mm}$ of the mm-VLBI instrument,
\begin{equation}
L^{\min}_{\nu} = \frac{L_{\nu}}{L_{(1+z)\nu}} 4 \pi D^2_L\frac{F^{\min}_{\nu}}{1+z}.
\end{equation}
where the first term is the K-correction, which accounts for a luminosity
difference of the quasar in the emitted ($(1+z)\nu$) and the observed ($\nu$)
bands. We use a fiducial value of $F^{\min}_{\nu} = 1$mJy for $\nu \sim 300$
GHz (mm wavelength), motivated by near-future capabilities of the Event
Horizon Telescope \citep{BrodLoebReid:2011A}.

To compute the K-correction, and to construct the mm-wavelength luminosity
function from a radio luminosity function, we assume that the spectral energy
distribution (SED) of Radio-loud AGN is a power law, $\nu L_{\nu} \approx
\nu^{0.9}$, over the frequency range $9 \lesssim \log_{10}(\nu/\rm{Hz}) \lesssim
12$ \citep{Elvis+1994}. This approximation is valid for LLAGN, as well as
regular radio-loud AGN, because the radio to millimeter portion of radio-loud
AGN spectra is similar for LLAGN and normal AGN \citep{Fern-Onti:LLSED+2012}.

To scale from $L_{mm}$ to a bolometric luminosity for LLAGN we use a median
bolometric correction from $2-10$ keV of $50$ \citep{Eracleous:LLAGN+2010}.
Because $\nu L_{\nu}$ in the millimeter is within a factor of a few of $\nu
L_{\nu}$ in the $2-10$ keV range \citep{Fern-Onti:LLSED+2012}, and because
there is a large scatter in the value of the bolometric correction in the
X-ray range, we adopt $L_{\rm{bol}} = 50 \nu_{mm} L_{mm}$.

By calculating $N_{\EHT}$ in this way, the mass distribution of binaries is
provided through the AGN luminosity function. We stress again that the
parameter $f_{\Edd}$, along with the accretion efficiency $\eta=0.1$ relates
the accretion rate on to the binary to the observed bolometric luminosity.
This is independent of $\mde$ above which parameterizes the
binary decay rate.

For each computation of the integrand in Eq. (\ref{Eq:NEHT}), corresponding to
a value of the bolometric luminosity, we draw a value of $\log_{10} f_{\Edd}$ from
the chosen Eddington fraction distribution. The value of $\log_{10} f_{\Edd}$ is
used to convert the luminosity for which the mmALF is evaluated to a binary
mass for which the probability $\mathcal{F}$ is evaluated. The probability
also depends on the size of the mm-wavelength emission region. This is
calculated from the luminosity and compared to the size of the binary
separation which depends on the binary mass through the value of the Eddington
ratio (Appendix \ref{S:mmSize}).

One caveat of our implementation is that the Eddington fraction distribution
is constructed as to match observations of low-luminosity, nearby AGN, and not
necessarily radio-loud AGN (\textit{i.e.}, we have assumed that the $f_{\Edd}$
distribution is the same for both radio-loud and radio-quiet populations.).
Meanwhile the mmALF we use is aimed at large redshifts for only radio-loud
AGN. The use of one with the other can cause extrapolation to very large BH
masses. This is because, at large $z$, only the most luminous AGN can be
observed. For a given Eddington fraction distribution, the most luminous AGN
must be powered by the most massive MBHs. Because the LLAGN Eddington
distribution samples very low values, the mapping from luminosity to binary
mass can result in very large values of the MBH mass, above $10^{10} \Msun$.
It is likely that such MBHs do not exist, but rather, that there would be a
preference for the Eddington distribution to sample larger values of
$f_{\Edd}$ at higher redshifts. To get around this artificial MBH mass
inflation, we simply make a cut in the mass distribution, so that,
effectively, low values of $f_{\Edd}$ are not sampled at high luminosities and
high redshifts. Essentially we are requiring that the LLAGN Eddington
distribution we have employed is correct for low redshifts where it is
derived, but then to extrapolate to high redshifts, we enforce that the
resulting MBH masses are consistent with observed maximum masses. We implement
this by multiplying the binary probability $\mathcal{F}$ by a factor
$\exp{\left[-(M/M_{\max})^4\right]}$. As a fiducial value we choose $M_{\max}
\sim 10^{10} \Msun$.

Choosing a maximum binary orbital period of $P_{\base}=10$ years and choosing
a minimum binary mass ratio of the entire MBHB population of
$q^{\Pmin}_s=0.01$, we are left with the free parameters $\theta_{\min}$,
$f_{**}$, $f_{\bin}$, $q^{\Vmin}_s$, $a_{\max}$, and $\mde$.
For mm VLBI, $\theta_{\min}$ is dominated primarily by
uncompensated propagation delays caused by the troposphere, rather than
contributions due to thermal noise that scale with the inverse of the
signal-to-noise ratio \citep{ReidHonma:2014,BrodLoebReid:2011A}. Hence we do
not adopt a signal-to-noise ratio-dependent resolution. Instead, we treat
$\theta_{\min}$ as a parameter that elucidates the gain in resolvable MBHBs
that can be achieved with better resolution limits. As motivated by the
gravitational wave background upper limits in the next section, we set
$f_{\bin}=0.05$. We set $f_{**} =1$ throughout, as the result scales linearly
with this factor. Then we determine the dependence of $N_{\EHT}$ on the turn-
on separation $a_{\max}$, minimum (symmetric) mass ratio of the resolvable
population $q^{\Vmin}_s$, and gas driven migration rate $\mde$ out to a given
redshift. The parameters of our model and their fiducial values are given in
Table \ref{Table:params} for reference.

\begin{table*}
%
%
\begin{tabular}{l|l|l|l|l}
  Parameter         & Meaning     & Fiducial Value & LLAGN Value  &   Notes \\
\hline
\hline 
Decay Model &  & \\
\hline
\hline 
$\mde$ \quad    & Gas supply rate in units of Eddington  & $1.0$  & "  & Fraction of supply rate controlling  \\
&&&& migration rate, not luminosity   \\
$a_{\max}$ \quad    & Separation where binary becomes active    & $0.1$ pc & "  & Needed because residence time in gas  \\
&&&& dominated regime is independent of $a$  \\
$q^{\Vmin}_s$ \quad    & Minimum binary symmetric mass ratio in VLBI sample   & $0.01$ & " & --  \\ 
$q^{\Pmin}_s$ \quad    & Minimum binary symmetric mass ratio of all MBHBs     & $0.01$ & " & --  \\ 
\\[-0.2cm]
\multicolumn{4}{|c|}{\it --- --- Additional decay-model parameters that are not varied --- --- }\\
\\[-0.2cm]
$f_{\Edd}$ \quad    & Fraction of Eddington luminosity     & $10^{-4.1}$ & Eq. (\ref{Eq:fEddDist}) & Relates $M$ to $L_{\rm{bol}}$  \\
$\eta$ \quad    & Accretion efficiency     & $0.1$ & "  & --  \\
$f_{\bin}$ \quad    & Fraction of AGN triggered by MBHBs     & $0.05$ & " & --  \\ 
$M_{\max}$ \quad    & Maximum allowed MBHB mass     & $10^{10}\Msun$ & " & See the end of \S \ref{S:Calc}  \\ 
$L_{\rm{bol}}/L_{mm}$ \quad    & Bolometric correction from millimeter     & $50$ & " & Estimate based on \cite{Eracleous:LLAGN+2010}  \\ 

\hline
\hline   
Observational &  & \\
\hline
\hline 
 $\theta_{\min}$    &  Minimum angular resolution        & $1 \rightarrow 30$ $\mu$as   & "   & \cite{BrodLoebReid:2011A}  \\
 $F_{\min}$    &  mm-Flux sensitivity        &  $1$ mJy &  "    &  \cite{BrodLoebReid:2011A} \\
 $P_{\base}$    &  Maximum observed binary period        &  $10$ years & "   & Detectable in human lifetime \\
$f_{**}$ \quad    & Fraction of MBHBs with \textit{both} BHs mm-bright    & $1.0$ & "  & --  \\ 
\end{tabular}
 %
 %
\caption{Parameters of the model and their fiducial values unless stated otherwise.}
\label{Table:params}
\end{table*}

\subsection{Gravitational Wave Background}
\label{S:GWB}

Before examining the population of resolvable MBHBs, we use the results of the
previous section to compute the contribution to the stochastic gravitational
wave background (GWB) of all MBHBs in this mock population and enforce
consistency with current limits from the PTAs. 

The frequency dependent, characteristic strain due to the gravitational wave
background is \citep[][]{Phinney:2001},
\begin{widetext}
\begin{equation}
h^2_c(f) = \frac{G}{c^2} \frac{4}{\pi f^2_r} \int^{\infty}_0 \int^{\infty}_0 \int^{1}_{q^{P\min}_s}{ \frac{d^3n}{dz dM dq} \frac{1}{1+z}  \frac{d E_{\rm{GW}}}{d \ln f_r} dq dM dz}.
\end{equation}
Here $f_r$ is the rest frame frequency of the GWs. Assuming circular binaries,
$f_r$ is equal to twice the Keplerian orbital period of the binary. The first
term under the integrand is the co-moving number density of inspiraling MBHBs
per redshift $z$, binary mass $M$, and mass ratio $q$. Continuing the assumption of
circular binary orbits, each MBHB emits GW energy per log frequency
\citep{SVC08},
\begin{eqnarray}
\frac{d E_{\rm{GW}}}{d \ln f_r} &=& \frac{d t_r}{d \ln f_r} \frac{32}{5} \frac{G^{7/3}}{c^5} \left(\pi M_c f_r \right)^{10/3} =   \frac{64}{15} \frac{G^{7/3}}{c^5} \left(\pi M_c f_r \right)^{10/3} t_{\rm{res}},
\end{eqnarray}
where we have rewritten the rest frame time per log frequency in terms of the
residence time of the binary at a given separation \citep[see,
\textit{e.g.},][]{KocsisSesana:2011}. This residence time is given by Eq.
(\ref{Eq:FGWGas}) for $a \leq a_{\max}$ and by the gravitational wave
residence time for $a > a_{\max}$. Here $M_c(M,q) \equiv M q^{3/5} /
(1+q)^{6/5}$ is the chirp mass of the binary and $q$ is related to the
symmetric mass ratio via the expression in the discussion below Eq.
(\ref{Eq:FGWGas}).

For the co-moving number density we use the luminosity function from the
previous section (see Appendix \ref{S:mmALF}), mapping luminosity to binary
mass via Eq. (\ref{Eq:MofL}). We further assume a flat distribution of mass
ratios from $q^{P\min}_s$ to $1$. Then the GWB strain becomes,
\begin{equation}
h^2_c(f) = \frac{G^{10/3}}{c^7} \frac{256}{15 \pi f^2_r} \int^{\infty}_0 \int^{\infty}_0{ \frac{d^2n}{dz dM} \left< \left(\pi M_c f_r \right)^{10/3} \frac{t_{\rm{res}}}{(1+z)} \right>_{q_s}  dz dM},
\end{equation}
\end{widetext}
where $\left< \cdot \right>_{q_s}$ denotes an average over the symmetric mass
ratio, $q_s$.

\begin{figure}
\begin{center}
\includegraphics[scale=0.42]{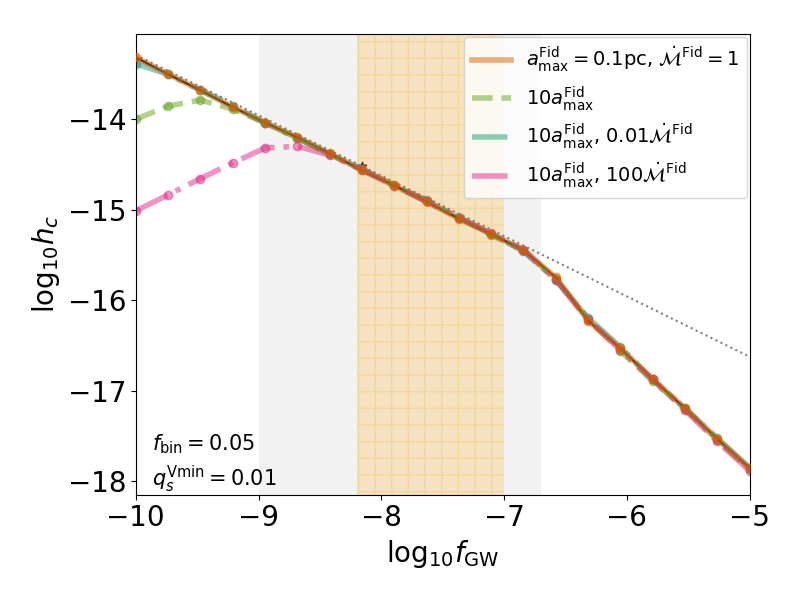}
\end{center}
\vspace{-10pt}
\caption{
The gravitational wave background of radio-loud MBHBs,
assuming that $5\%$ of radio-loud AGN harbor a MBHB. The dashed gray line is
the GW background due to gravitational wave-driven decay alone and with
amplitude calibrated by PTA upper limits. The gray shaded region is
the approximate PTA frequency range, and the hashed orange region is the range
of inspiral frequencies where MBHBs could be resolvable by millimeter VLBI.
}
\label{Fig:GWB}
\end{figure}

The GWB characteristic strain associated with the radio-loud MBHBs is plotted
versus GW frequency in Figure \ref{Fig:GWB} for four different sets of
parameters dictating the gas-driven decay, and assuming the power law plus
Gaussian Eddington ratio distribution described in the previous section. The
dashed-gray line plots the approximate, current PTA upper limits assuming GW
decay alone and scaled to the currently most stringent PTA upper limits
at a single frequency \citep{SesanaHK:2017, Shannon:2015}. The hatched orange
region represents the approximate range of GW frequencies for which 
mm-VLBI-resolvable MBHBs will reside, while the gray shaded region represents the
approximate PTA frequency range.

For the fiducial parameters (solid-orange line), the GWB is dominated by
gravitational decay as can be seen by the $f^{-2/3}_{\GW}$ decline in the
background at low frequencies. The increase in slope above $\sim 10^{-7}$Hz is
due to the removal of MBHs that would be at separations smaller than that of
the last stable orbit corresponding to the binary mass ($a=6GM/c^2$).

Increasing both $\mde$ and $a_{\max}$ increases the impact of the gas-driven
decay on the GWB at low frequencies, in agreement with the
studies of \cite{Kelley+2017} and \cite{KocsisSesana:2011}. This can be seen by
the turn over of the green dashed and pink dot-dashed lines in Figure
\ref{Fig:GWB} at low frequencies. The gas moves the binary through this larger
separation regime more quickly than  GW-driven decay alone.

We find that the population of radio-loud AGN must have a MBHB fraction of
$\lesssim 0.05$ for consistency of our model with the upper limits on the PTA
GWB. Hence we use that $f_{\rm{bin}}= 0.05$ as our fiducial MBHB fraction
throughout. 

We note that there are many studies of the GWB from model
MBHB populations that find agreement with the PTA upper limits, but with a
less stringent constraint on the fraction of MBHB harboring AGN
\citep[see][and references therein]{KocsisSesana:2011, KelleyBH:2017, Kelley+2017,
Mingarelli+2017}. The stringency of the limit derived here is the
inclusion of LLAGN via the Eddington distribution plotted in Figure
\ref{Fig:GWB}; the many AGN at low Eddington ratios implies more MBHBs at
higher masses. This is evidenced in the non-trivial dependence of the GWB on
the choice of Eddington fraction distribution. If, for example, we choose
$x_{\min}=-4$ instead of the fiducial $x_{\min}=-5.5$ in Eq.
(\ref{Eq:fEddDist}), then the inferred GWB strain is an order of magnitude
smaller, and hence $f_{\bin}$ is not constrained at all. We further note that this GWB is
calculated using the radio-loud luminosity function, which accounts for $10\%$
of the AGN. However, as discussed above, our value of
$f=0.05$ is already quite conservative and dependent on the Eddington
distribution fraction. Hence, we simply take a fiducial value of $f=0.05$
which is the minimal constraint taking into account all of the MBHBs
represented in our model and move
forward within our toy-model, using the already motivated choice of Eddington
fraction distribution. We leave exploration of this distribution, and
consequences for the GWB, to future work.

\section{Results}
\label{S:Results}

\subsection{Number of resolvable MBHBS: redshift distribution and dependence on angular resolution}
\label{S:Results:a}

\begin{figure*}
\begin{center}$
\begin{array}{c c}
\includegraphics[scale=0.44]{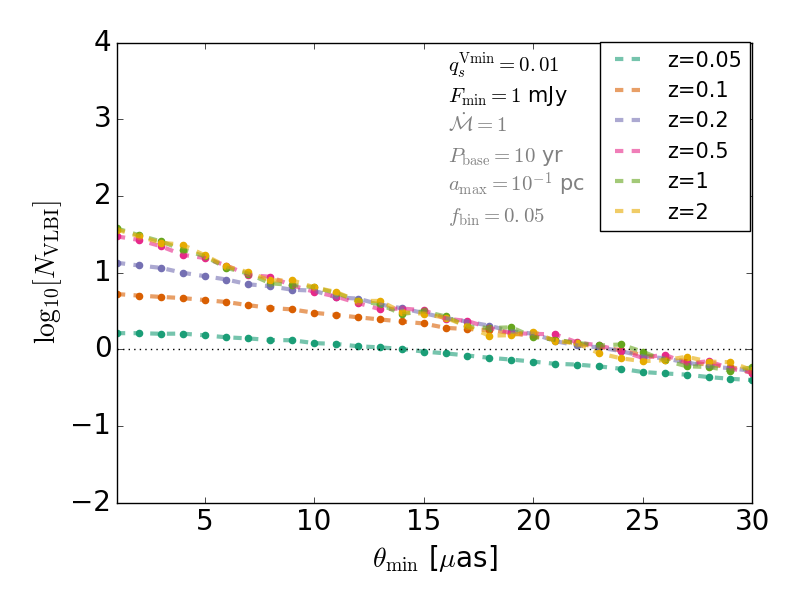} &
\includegraphics[scale=0.44]{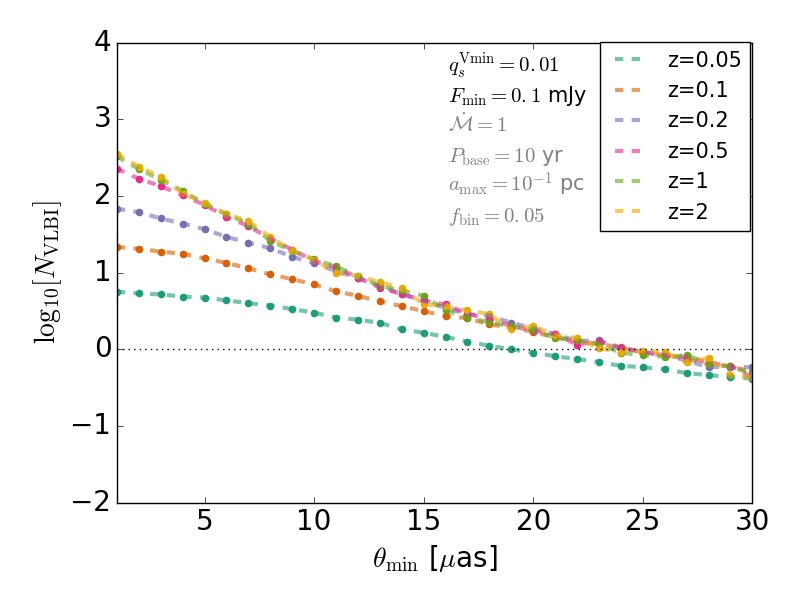} \\
\includegraphics[scale=0.44]{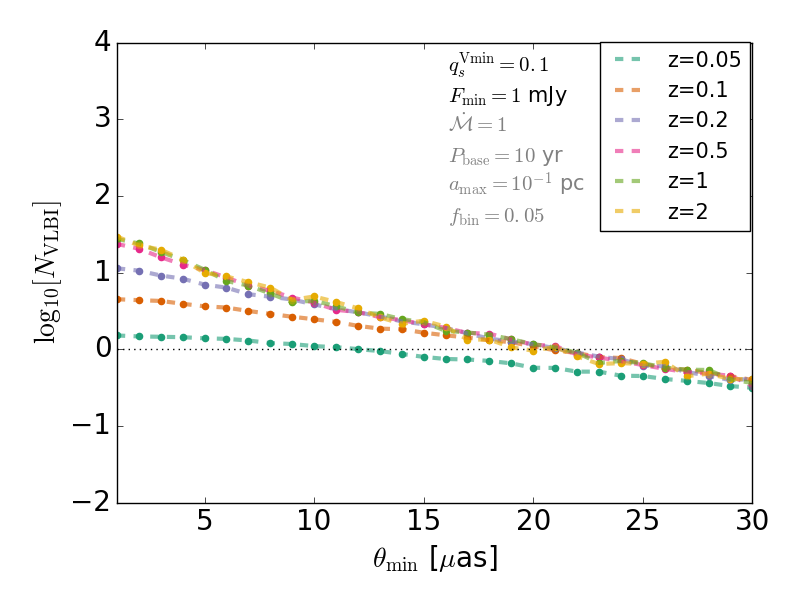} &
\includegraphics[scale=0.44]{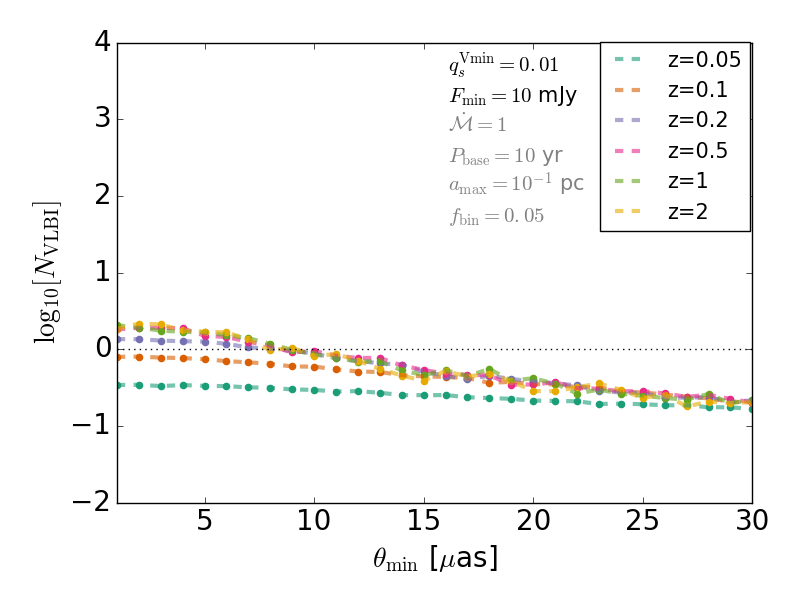} \\
\includegraphics[scale=0.44]{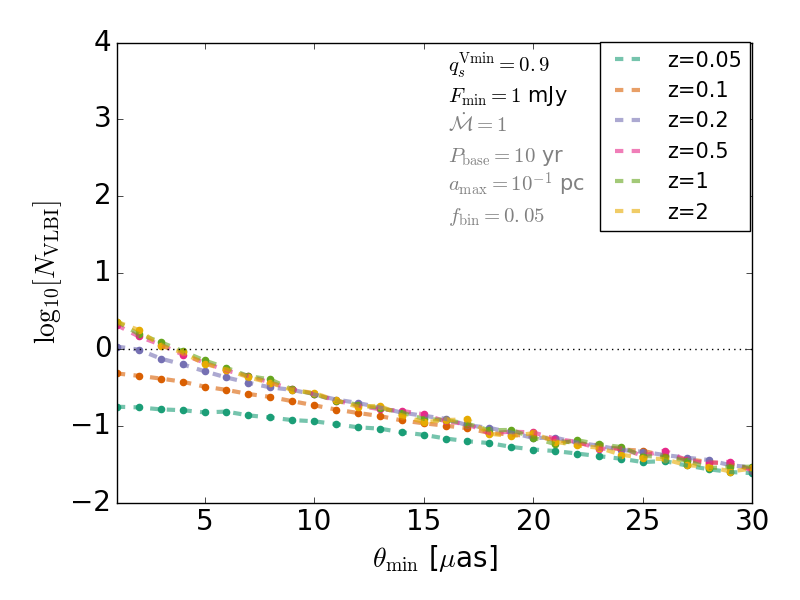} &
\includegraphics[scale=0.322]{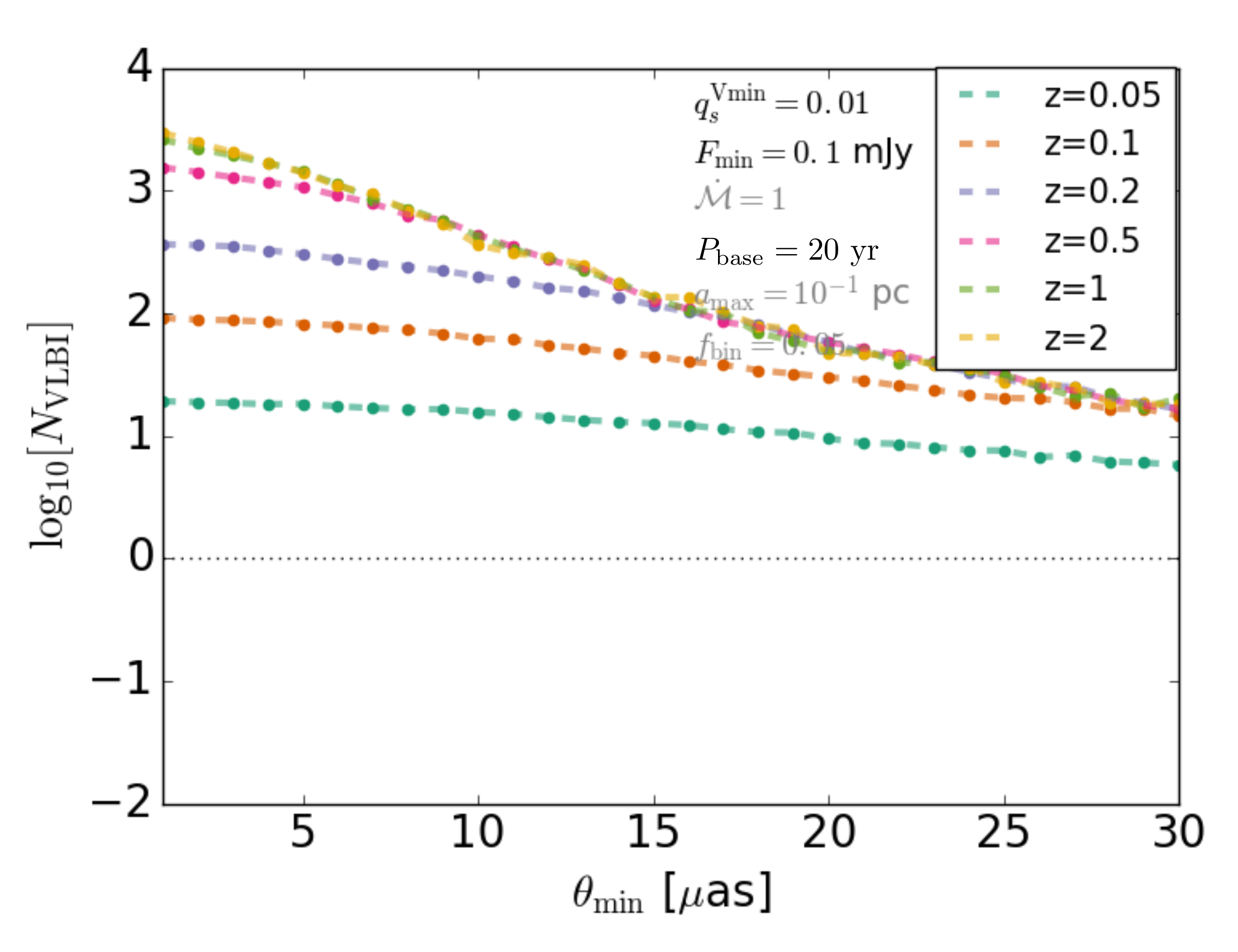} 
\end{array}$
\end{center}
\vspace{-10pt}
\caption{
The number of resolvable MBHBs over the entire sky as a function of minimum
instrument angular resolution $\theta_{\min}$ for different maximum redshifts
(labeled). The left column fixes the flux sensitivity to the fiducial  value
but increases the minimum binary mass ratio. The top right panel enhances the
instrument flux sensitivity, the middle right panel worsens the instrument
flux sensitivity, and the bottom right panel shows the gain due to increasing
the maximum baseline period at the best case flux sensitivity.
}
\label{Fig:N_vs_thmn}
\end{figure*}

We begin by presenting the total number of resolvable MBHBs
for which an entire orbit can be tracked with VLBI. Figure
\ref{Fig:N_vs_thmn} explores the dependence of the total number of resolvable
binaries on maximum redshift and minimum instrument angular resolution. The
binary fraction in AGN is assumed to be $f_{\bin} = 0.05$, as constrained in
\S \ref{S:GWB}. The left column varies the minimum binary mass ratio from
$q^{\Vmin}_s=0.1$ to $q^{\Vmin}_s=0.9$ for $F_{\min} = 1$mJy. The top two
panels in the right column consider a lower and higher minimum flux
sensitivity, at a fixed minimum mass ratio of $q^{\Vmin}_s=0.01$. The bottom
right panel considers a larger maximum binary period at the optimal minimum
mass ratio and flux sensitivity.

In the fiducial case (top left panel), which assumes no mass ratio preference
for the resolvable binaries ($q^{\Vmin}_s = 0.01$) and a flux sensitivity of
$F_{\min}=1$mJy, we could resolve binaries with angular separations as large
as $\sim25\mu$as. This is within the diffraction limit of present day mm-VLBI.
At a best case angular resolution of $\theta_{\min} = 1 \mu$as, a few tens of
MBHBs would be resolvable with mm-VLBI. These resolvable MBHBs lie between
redshift $0.05$ and $0.5$.

The left column of Figure \ref{Fig:N_vs_thmn} shows that in order to resolve at
least one MBHB, one must have $q^{\Vmin}_s \lesssim 0.9$, at the best case
angular resolution and fiducial flux sensitivity. This implies that a resolvable
population, for the fiducial binary-decay model, can not consist of preferentially
equal mass binaries (but it can contain equal mass binaries). 

The top right panel of Figure \ref{Fig:N_vs_thmn} shows that an order of magnitude
improvement in flux sensitivity provides an order of magnitude increase in
$N_{\EHT}$. This improvement in sensitivity also increases the redshift range
out to $z\sim1.0$. The improved sensitivity does not increase the range of
$\theta_{\min}$ at which resolvable MBHBs can be found. The middle right panel
of Figure \ref{Fig:N_vs_thmn} shows that an order of magnitude worse flux sensitivity
provides an order of magnitude decrease in $N_{\EHT}$. The middle right panel
also shows that for $N_{\EHT} \gtrsim 1$, one must have $F_{\min} \lesssim
10$mJy for the best case $q^{\Vmin}_s$.

Finally, the bottom right panel of Figure \ref{Fig:N_vs_thmn} considers a best case
scenario: including MBHBs with orbital periods up to $20$ years, an optimal
flux sensitivity of $F_{\min} = 0.1$mJy, and otherwise fiducial model
parameters. We see that the number of resolvable MBHBs can be increased to a
few thousand at the best case angular resolution. Including longer period
binaries also yields a population of large angular separation, $\gtrsim 30
\mu$as separation binaries.

\subsection{Visualization of result}
\label{S:Results:b}

We now investigate the dependence of our results presented in Figure
\ref{Fig:N_vs_thmn} upon the parameters of our model. This allows us to
determine the demographic of MBHBs that can be tracked with VLBI, and it also
allows us to determine the range of results that our model could produce.

\begin{figure*}
\begin{center}$
\begin{array}{c c}
\includegraphics[scale=0.46]{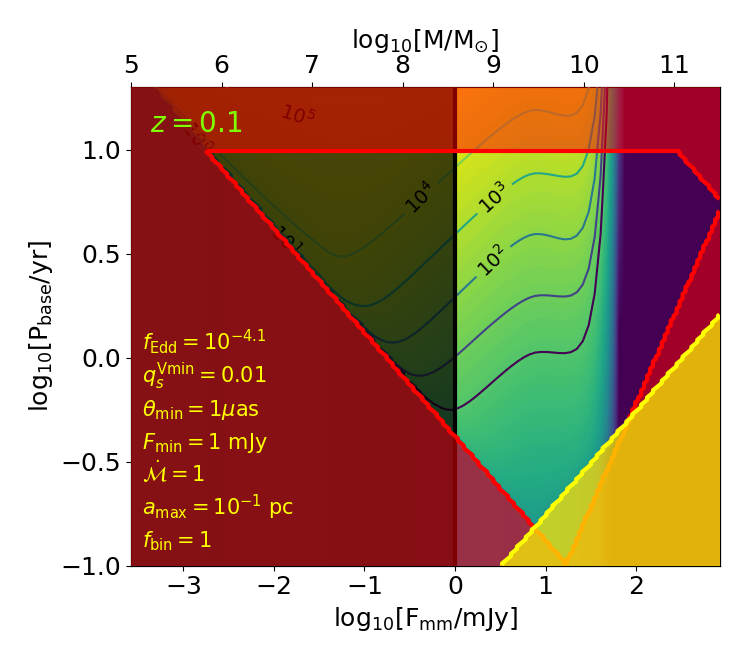} &
\includegraphics[scale=0.46]{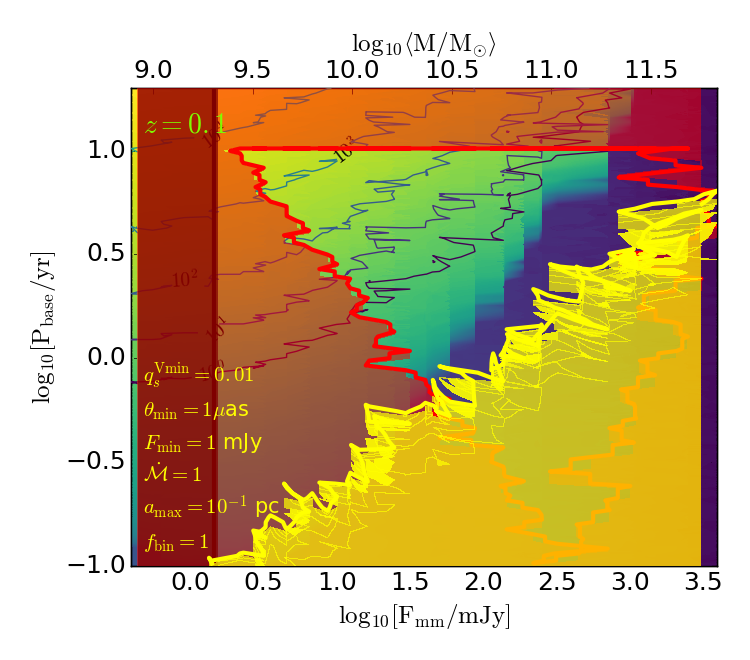} \vspace{-0.3cm}\\ 
\includegraphics[scale=0.27]{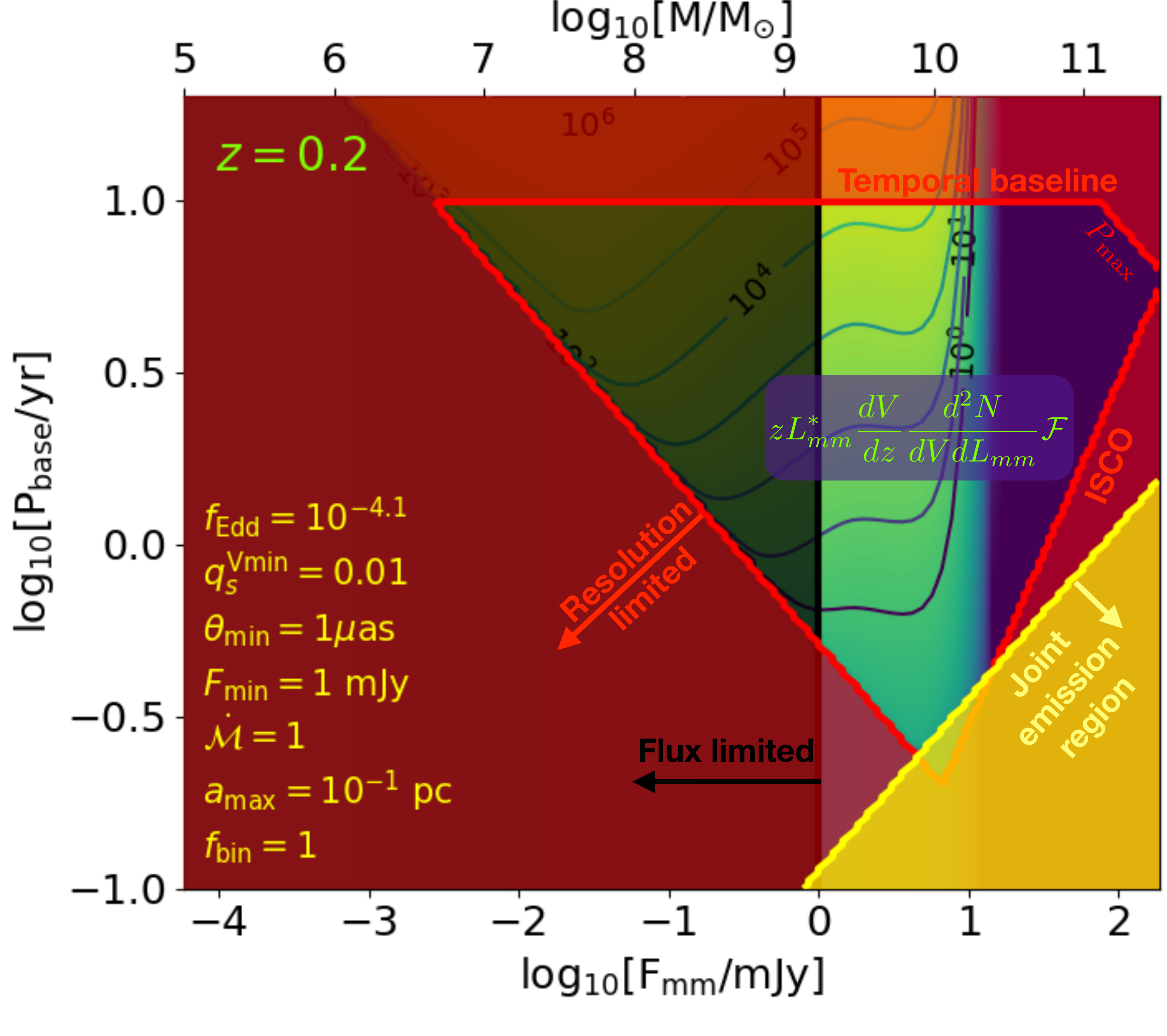}  &
\includegraphics[scale=0.46]{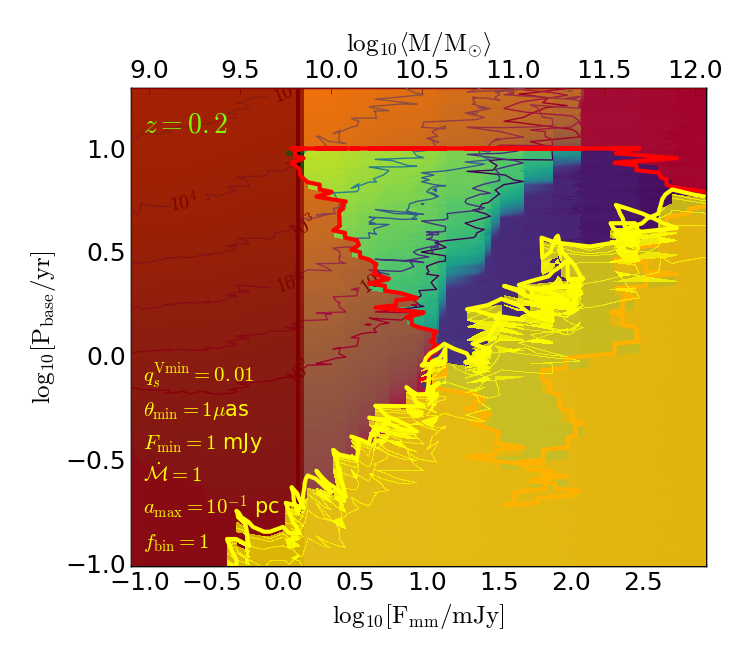} \vspace{-0.3cm} \\
\includegraphics[scale=0.46]{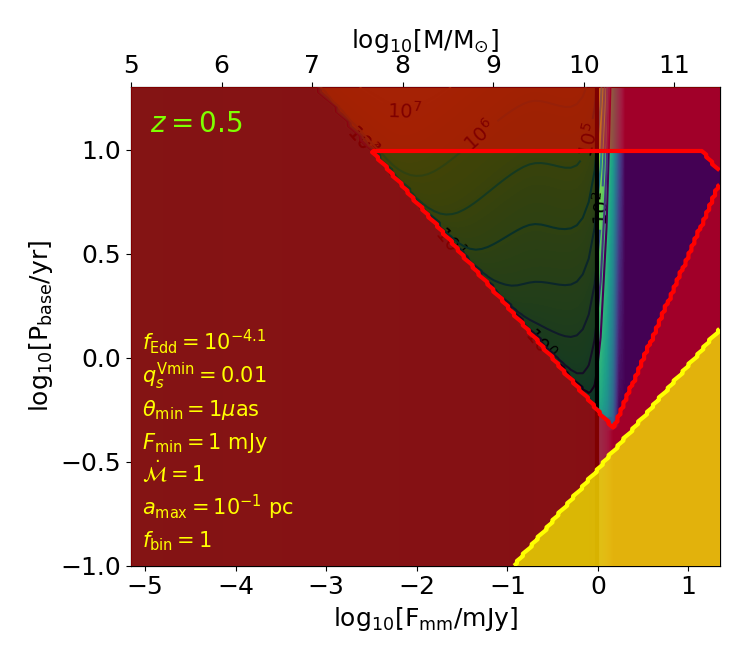} &
\includegraphics[scale=0.46]{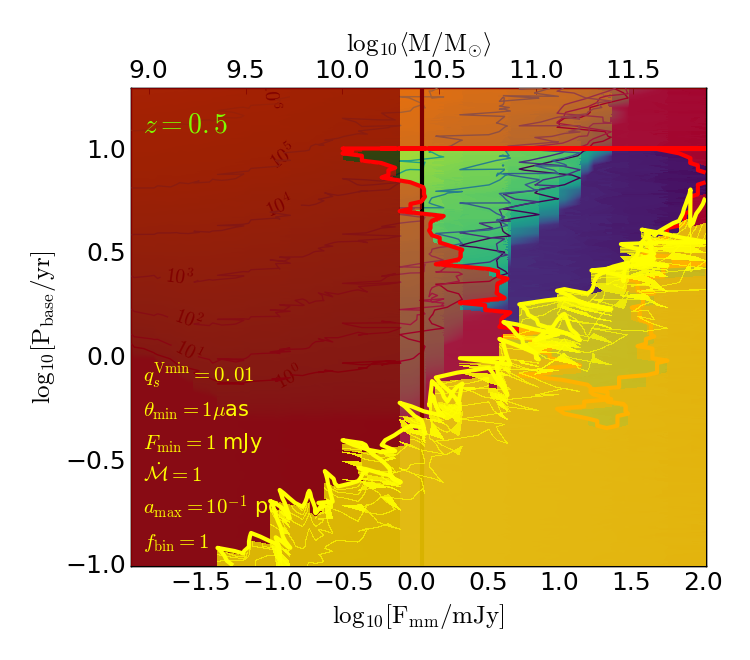}
\end{array}$
\end{center}
\vspace{-10pt}
\caption{
Visualization of the integral of Eq. (\ref{Eq:NEHT}). The
left middle panel is labeled as a guide to the the different regions plotted
in each panel. The left column assumes a delta function value of the
Eddington ratio, $f_{\Edd} = 10^{-4.1}$. The right column assumes a power law
plus Gaussian distribution of $f_{\Edd}$ (Eq. \ref{Eq:fEddDist}) and averages
over 200 $f_{\Edd}$ draws. From top to bottom, each row is for redshift
$z=0.1$, $0.2$, and $0.5$. The labeled contours, shaded
from chartreuse to purple, are a scaling of the integrand in Eq.
(\ref{Eq:NEHT}), with chartreuse indicating many MBHBs and purple indicating
none. The differently colored shaded regions, overplotted on these contours, denote where our criteria for a
MBHB to be a VLBI tracking target are not met. Hence, the trapezoid-shaped
window in the center of each panel delineates the region of parameter space
where MBHBs can be resolved and tracked along their orbits with VLBI. The
\textbf{darker-shaded regions}, bounded on the right by the thick black line,
are where emission is too dim to detect, given the labeled minimum flux
sensitivity. The \textcolor{red}{red-shaded regions}, external to the thick
red lines, are where the binary is not resolvable either because it is below
the minimum resolvable separation, above the maximum baseline orbital period,
or at merger. The \textcolor{Dandelion}{yellow-shaded regions} are where the
mm-emission region is larger than the binary separation.
}
\label{Fig:VisInt}
\end{figure*}

We begin by visualizing the MBHB demographics that contribute to the integral
in Eq. (\ref{Eq:NEHT}). Each panel of Figure \ref{Fig:VisInt} plots contours
of the integrand of Eq. (\ref{Eq:NEHT}) as a function of maximum binary
orbital period and observed mm-wavelength flux. We scale the integrand  by $4
\pi z L_{\rm{mm}}$, taking a representative value of $L^*_{mm} =
10^{44}/\nu_{mm} \sim 3 \times 10^{32}$erg s$^{-1}$ Hz$^{-1}$ to correspond
roughly to the cumulative value of the integral at that point.
Contours of this scaling of the integrand of Eq.
(\ref{Eq:NEHT}) are colored chartreuse to purple, denoting many MBHBs and zero
MBHBs respectively. On top of the chartreuse-purple contours we shade regions
with different colors corresponding to the regions where our criteria for a
resolvable orbit break down.  That is, the overplotted-shaded regions
delineate the space of resolvable (unshaded) and non-resolvable (shaded)
binaries. The left-middle panel labels these regions. The resulting,
trapezoid-shaped window at the center of each panel, which contains only the
chartreuse-purple contours, frames the parameter space of MBHBs that are
resolvable and have orbits that can be tracked in a human lifetime.

On the top horizontal axis of each panel in Figure \ref{Fig:VisInt}, we relate
the observed flux (bottom axis) to the binary total mass via the Eddington
ratio. The left column of Figure \ref{Fig:VisInt} assumes a delta function
value of the Eddington ratio, $f_{\Edd} = 10^{-4.1}$, that corresponds to the
expectation value of the distribution of Eq. (\ref{Eq:fEddDist}). The right
column samples from the full Eddington ratio distribution. Because the latter
case requires a random draw of the Eddington ratio for each point in 
period-flux space, we generate 200 realizations and plot the average. In the right
panels we also plot the average binary mass that corresponds to the flux on
the lower horizontal axis (using the same 200 $f_{\Edd}$ draws). In
each panel we record the fixed parameters in the bottom left.

The darkest regions to the left of the thick black lines
represent binaries that are dimmer than the limiting flux $F^{\min}_{mm}$. For
the fixed $f_{\Edd}=10^{-4.1}$ Eddington ratio case (left column of Figure
\ref{Fig:VisInt}), at $z=0.1$, MBHBs with total mass $\gtrsim10^{8.5} \Msun$
are detectable at a limiting flux of $F^{\min}_{mm}=1$mJy. By $z=0.5$, there
are no more MBHBs below $\sim10^{10} \Msun$ that are detectable above $1$mJy.

In the right column, this darker-shaded region corresponds to the same flux limit
as the left column, but this corresponds to a different binary mass via the
Eddington ratio distribution in Eq. (\ref{Eq:fEddDist}). In this case, using
the power law plus Gaussian distribution, the same fluxes correspond to higher
mass binaries. At $z=0.1$, MBHBs with total mass of $\gtrsim10^{9} \Msun$,
on average, are detectable at a limiting flux of $F^{\min}_{mm}=1$mJy. By
$z=0.5$, only rare MBHBs with mass $\gtrsim10^{10.5} \Msun$ can be bright
enough to exceed the flux limit. Hence, binaries can be resolved out to
$z\sim0.5$, limited by the minimum detectable mm-wavelength flux. These
binaries will preferentially be more massive at higher redshifts.

Note that in the right columns, contours for the scaled integrand of Eq.
(\ref{Eq:NEHT}) do not plunge abruptly at the maximum cut-off mass as they
do in the left panels. This is due to the averaging of $M$ for many draws of
$f_{\Edd}$.

The red regions, lying outside of the thick red trapezoid, represent the space
of restricted binary orbital parameters. Below the bottom-left red line, for
smaller masses and periods, the binary is too compact to be resolvable at the
quoted redshift and minimum instrument angular resolution (this being the
$P_{\min}$ limit of integration in Eq. \ref{Eq:NEHT}). The bottom-right red
line is the orbital period at $a=6GM/c^2$, which we label ISCO. The top,
horizontal red line is the imposed maximum observed orbital period
($P_{\rm{base}}$). To the far top right, the small restricted region at the
longest periods and largest binary masses is where the period at turn-on
separation $a_{\max}$ is shorter than the maximum baseline period (see
definition of $P_{\rm{hi}}$ in Eq. \ref{Eq:NEHT}).

In both left and right columns, it is evident that the trapezoid-shaped window
defined by the red regions gets smaller and moves to lower fluxes at higher
redshifts. Both behaviors are a combination of three effects: i) the minimum
angular resolution corresponds to a larger physical binary separation at
greater distances, \footnote{This is the case at the redshifts of interest
here. For $z\gtrsim1.6$, this relation reverses due to cosmology.} ii) higher
flux systems do not exist at larger distances because of the limiting
intrinsic luminosity set by the maximum binary mass and the Eddington ratio
distribution, and iii) at larger distances, higher flux systems correspond to
more massive binaries; for a fixed binary period these systems have smaller
binary separations and are less likely to lie above the limiting angular
resolution.

The yellow region, in the bottom right of each panel,
represents where the mm-wavelength emission region is larger than the binary
separation. This occurs for the closest separation (shortest orbital period)
and brightest systems because brighter emission corresponds to a larger
photosphere (see Appendix \ref{S:mmSize}). For the low Eddington ratios chosen
here, only the more rare, short period, high mass binaries are obscured by a
large photosphere. This is of course the reason that we consider LLAGN for
this study. Figure \ref{Fig:NEHT_amx_fEdd} shows that for Eddington ratios
larger than $\sim 0.1$ (and even lower for worse angular resolution limits)
obscuration by a large mm-photosphere engulfs what would otherwise be a
resolvable binary.

Summarizing Figure \ref{Fig:VisInt}, we find that:
\begin{itemize}
\item The number of resolvable MBHBs becomes flux limited at $z \gtrsim 0.5$ (for
$F^{\min}_{mm}=1$mJy).
\item 
MBHBs with total mass of $10^7-10^8 \Msun$ and with the
longest orbital periods contribute the most to $N_{\EHT}$. However, the
relevant flux sensitivity results in a preference for the brightest, highest mass
systems. Increasing $P_{\base}$ always yields more resolvable MBHBs.
\item The most massive and shortest period binaries are obscured by a large 
mm-wavelength emission region.
\item Lower mass and shorter period binaries are excluded by angular
resolution requirements.
\item For the small Eddington ratios considered here, the mm-emission region size
constraints do not exclude much of the parameter space. Size constraints do
become important, however, for Eddington ratios closer to unity (see Figure
\ref{Fig:NEHT_amx_fEdd} below).
\end{itemize}

\subsection{Parameter dependencies}
\label{S:Pdeps}

Now that we have explored the make up of MBHBs that contribute to the integral
in Eq. (\ref{Eq:NEHT}), we explore the dependence of the result, $N_{\EHT}$,
on the major model parameters. The most important parameters controlling the
binary decay model are the turn-on separation $a_{\max}$, the minimum binary
mass ratio $q^{\Vmin}_s$ and the gas driven decay rate $\mde$. In Figure
\ref{Fig:NEHT_amx_eps} we explore the dependence of $N_{\EHT}$ on these
parameters for three values of the minimum angular resolution. On the left
vertical-axis of each panel we plot the angular scale corresponding to
$a_{\max}$ at $z=0.2$. In each panel, we record the fixed parameters in the
top left. We graphically mark the fiducial values of the parameters being
varied by green lines intersecting at a green point. The white lines are
contours of constant residence time in units of years. These are drawn for a representative
binary with $M=10^9 \Msun$ and $q_s=0.1$ (left column) or $M=10^9 \Msun$ and
$\mde=1$ (right column) and labeled in units of years.

To demonstrate the full parameter dependencies of $N_{\EHT}$, we draw contours
of $N_{\EHT}$ in Figure \ref{Fig:NEHT_amx_eps} assuming a value of
$f_{\bin}=1$. Note, however, that the gravitational wave analysis of the
previous section requires $f_{\bin} \lesssim 0.05$. Hence, Figure
\ref{Fig:NEHT_amx_eps} shows that, for minimum resolutions ranging from $1$ to
$20 \mu$as, the binary decay model predicts a total of $N_{\EHT} \sim f_{**}
f_{\bin} 10^3$ resolvable binaries at the fiducial parameter values of
$F^{\min}_{mm} = 1$mJy, $a_{\max} \sim 0.1$ pc, $q^{\Vmin}_s=0.01$, and $\mde
\sim 1.0$. For the range of plausible parameter space, and relaxing the GWB
limit, this number can range up to $\sim 10^5$. Next we discuss the dependence
of $N_{\EHT}$ on these parameter values.

\begin{itemize}

\item $\boldsymbol{a_{\max}}$ \textbf{dependence:} Above and below $a_{\max}
\sim 10^{-1.75}$ pc, $N_{\EHT}$ decreases. For $a_{\max} \gtrsim 10^{-1.75}$
pc this results from the assumption that $a_{\max}$ corresponds to where the
binary becomes bright and, simultaneously, where  gas-driven orbital decay
begins. For larger values of $a_{\max}$, the
range of binary separations over which the binary is bright becomes larger
while the range of binary separations over which it is resolvable stays the
same. Put another way, for larger $a_{\max}$, the space of possible binary
parameters for which the binary is bright increases in size while the target
range does not, and so the probability that any bright MBHB system is in the
resolvable range decreases.

For $a_{\max} \lesssim 10^{-1.75}$pc, the space of possible binary parameters
over which the binary is resolvable decreases because the binaries with the
longest allowed periods ($P\rightarrow P_{\base} (1+z)$) have separations
larger than $a_{\max}$ and hence are not bright all of their resolvable
lifetime. $N_{\EHT}$ drops to zero at the value of $a_{\max}$ that falls below
the minimum resolvable angular resolution at a given redshift. This can be
seen by comparing the location of the small-$a_{\max}$ cutoff of $N_{\EHT}$
between the $\theta_{\min}=1$, $10$, and $20 \mu$as panels in Figure
\ref{Fig:NEHT_amx_eps}.

Physically motivated values of $a_{\max}$ are suggested by the red shaded
regions in Figure \ref{Fig:NEHT_amx_eps}.  This region is the range of (binary
mass dependent) outer radii of a gravitationally stable gas disk
\citep{Goodman:2003, HKM09}.  To illustrate this range we choose a binary mass
of $10^9 \Msun$ (motivated by Figure \ref{Fig:LLAGN_PvM} below) and compute
the range of gravitationally stable outer disk radii assuming electron
scattering and  free-free absorption dominated opacity.

However, because the bright lifetime of an AGN is not
necessarily determined by the size of a Toomre-stable disk, we leave
$a_{\max}$ as a free parameter. We choose a fiducial value of
$a_{\max}=0.1$pc to be consistent with these gravitationally stable disks, but
also to correspond to a value near the peak of the $N_{\EHT}$ distribution. It
is encouraging that the most optimistic predictions for $N_{\EHT}$ lie within
this region. We do not choose a smaller $a_{\max}$, because of possible
tension with the sub-pc separations of known MBHB candidates
\citep[\textit{e.g.}][]{Graham+2015b, Charisi+2016}. Smaller $a_{\max}$, and
hence larger inferred values of $N_{\EHT}$, are of course not ruled out.

\item{$\boldsymbol{q^{\Vmin}_s}$ \textbf{dependence:}} In the left column of
Figure \ref{Fig:NEHT_amx_eps}, we plot contours of $N_{\EHT}$ in $a_{\max}$
vs. $q^{\Vmin}_s$ space. That is, we assume a minimum mass ratio
$q^{\Pmin}_s=0.01$ of the population of all MBHBs and vary the minimum mass
ratio of resolvable MBHBs, $q^{\Vmin}_s$. The utility of the $q^{\Vmin}_s$
parameter is to elucidate which mass ratios are contributing to the resolvable
population. It is also useful if we wish to restrict the population of
resolvable MBHBs to be only those with near unity mass ratios, to increase the
probability that both will be bright. The left panels of Figure
\ref{Fig:NEHT_amx_eps} clearly show that the value of $N_{\EHT}$ only begins
to strongly depend on $q^{\Vmin}_s$ when the latter approaches unity, at which
point $N_{\EHT}$ drops to zero (at $q_s\sim0.7$ or $q\sim0.3$). This is
simply a result of positing a flat distribution of binary mass ratios.

\item $\boldsymbol{\mde}$ \textbf{dependence:} In the right column of Figure
\ref{Fig:NEHT_amx_eps} we plot contours of $N_{\EHT}$ in $a_{\max}$ vs. $\mde$
space. To understand the dependence of $N_{\EHT}$ on $\mde$, we first consider
the purely  \textit{gas-driven} scenario. In this case, $N_{\EHT}$ does
\textit{not} depend on $\mde$, rather it sets the average active lifetime of
the MBHB indicated by the white contours in Figure
\ref{Fig:NEHT_amx_eps}. 
The steep turn-around in these white contours is where the binary changes from
gas-driven to GW-driven orbital decay. In the gas-driven regime, there is a
decrease in MBHB lifetime with increasing $\mde$ as expected, but this is not
reflected in the contours of $N_{\EHT}$ because the binary decreases its time
spent in the resolvable regime proportionally to its total lifetime. This is
essentially a result of the duty-cycle argument that we use in this
computation, \textit{i.e.}, our assertion that a fraction $f_{\bin}$ of AGN
are MBHBs. Reassuringly, our total MBHB lifetimes are consistent or slightly
lower than observationally constrained AGN lifetime \citep{PMartini:2004}.
We note, however, that the bright lifetime of an LLAGN may
be longer ($\sim10-100\times$) than the bright Eddington-limited quasar
lifetime \citep{HopkinsLidz+2007}. Hence, choices of $\mde$ (and $a_{\max}$)
that lead to $\sim$Gyr bright binary residence times may be relevant for the
LLAGN.

When including GW emission, $N_{\EHT}$ does depend on $\mde$ because it sets the
binary separation where GW-driven decay takes over. The consequence being that
larger values of $\mde$ correspond to binaries that spend relatively more time
in the gas driven regime than in the more short-lived GW-driven regime. This
affects the value of $N_{\EHT}$ if the transition separation $a_{\trans}$ falls
within $a_{\max}$. The wider the range of separations that the binary spends
in the gas driven stage, and above the minimum separation $a_{\min} =
\theta_{\min} D_A(z)$, the more resolvable binaries we expect to find.

This explains why $N_{\EHT}$ is constant below values of $a_{\max}$ on a line
parallel to the line connecting the elbows of the white contours, the
transition from gas to GW driven decay. Below this line $a_{\max} <
a_{\trans}$, and the binary is already in the GW-driven regime when it enters
the disk and becomes bright, so gas does not affect the decay. That $N_{\EHT}$
obtains an $\mde$ dependence approximately above the line connecting the
elbows of the white, $M=10^9 \Msun$ contours, means that these most-massive
MBHBs dominate the resolvable population. We show that this is indeed the case
in Figure \ref{Fig:LLAGN_PvM}.

The increasing dependence of $N_{\EHT}$ on $\mde$ with larger $\mde$ stems
from the fact that larger $\mde$ implies a smaller $a_{\trans}$ and hence
causes the binary to spend more of its resolvable lifetime in the gas
dominated stage. A similar increasing importance of gas effects for larger
$a_{\max}$ and smaller $a_{\trans}$ (larger $\mde$), can also been seen in the
different GWB realizations of Figure \ref{Fig:GWB}.

\begin{figure*}
\begin{center}$
\begin{array}{c c}
\includegraphics[scale=0.46]{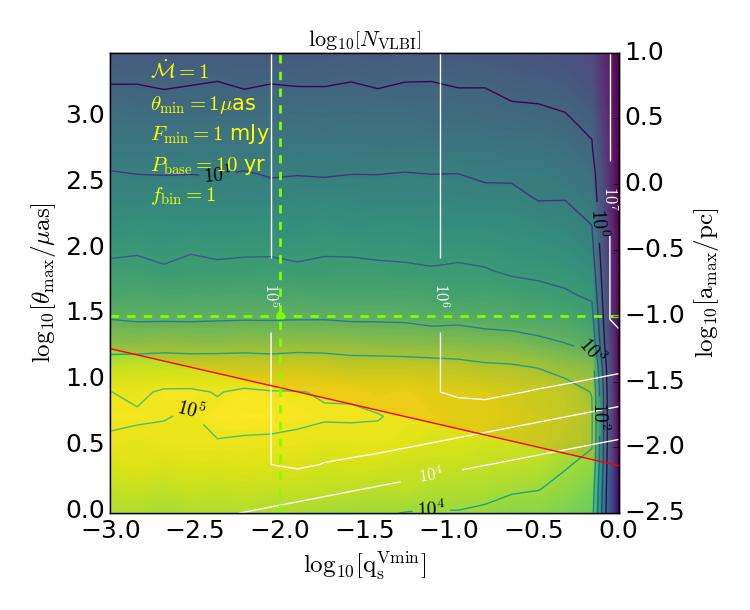} &
\includegraphics[scale=0.46]{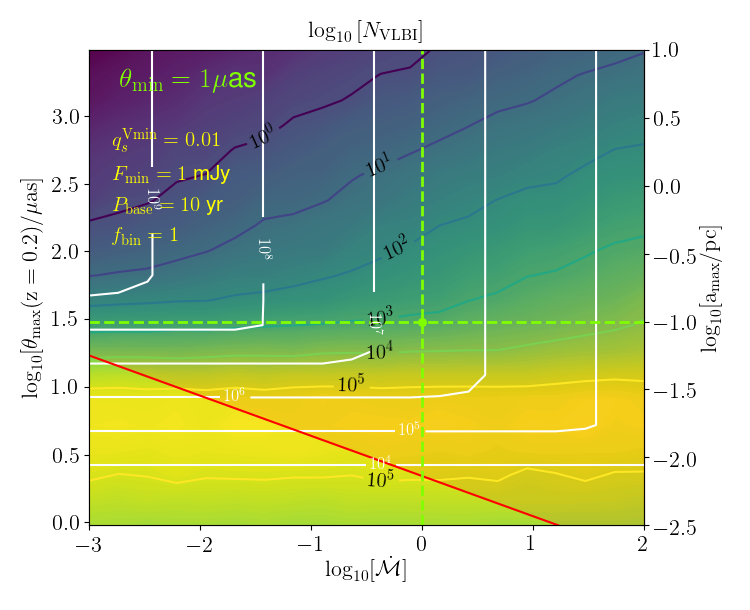} \vspace{-0.4cm} \\
\includegraphics[scale=0.46]{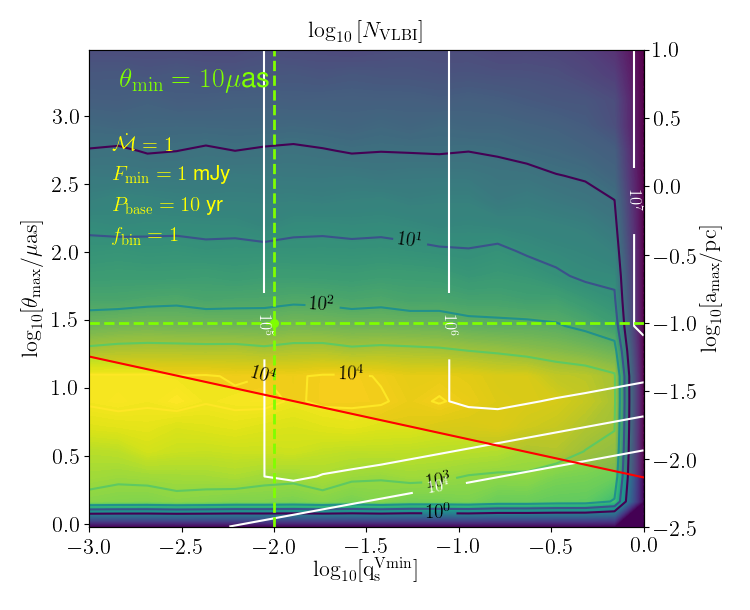} &
\includegraphics[scale=0.46]{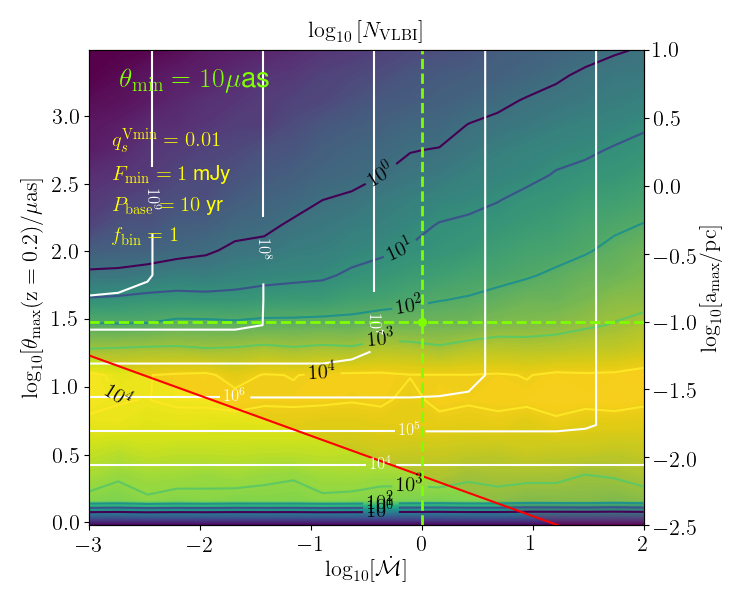} \vspace{-0.4cm} \\
\includegraphics[scale=0.46]{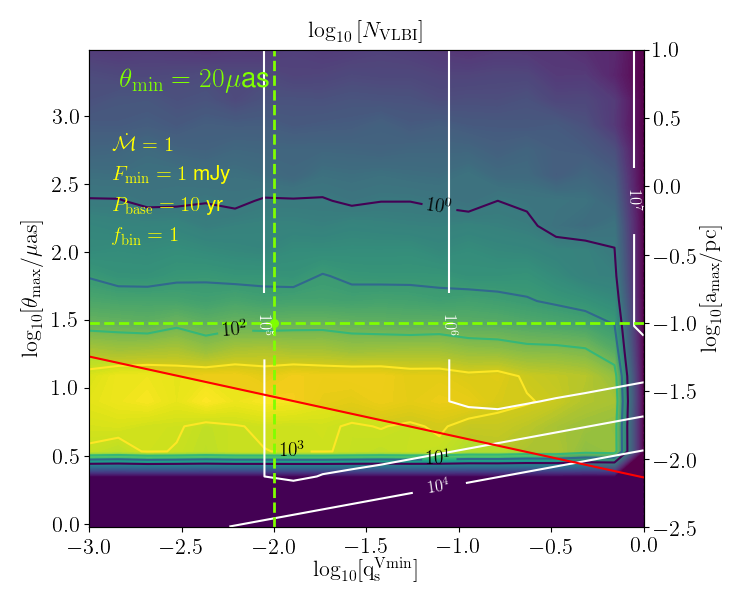} &
\includegraphics[scale=0.46]{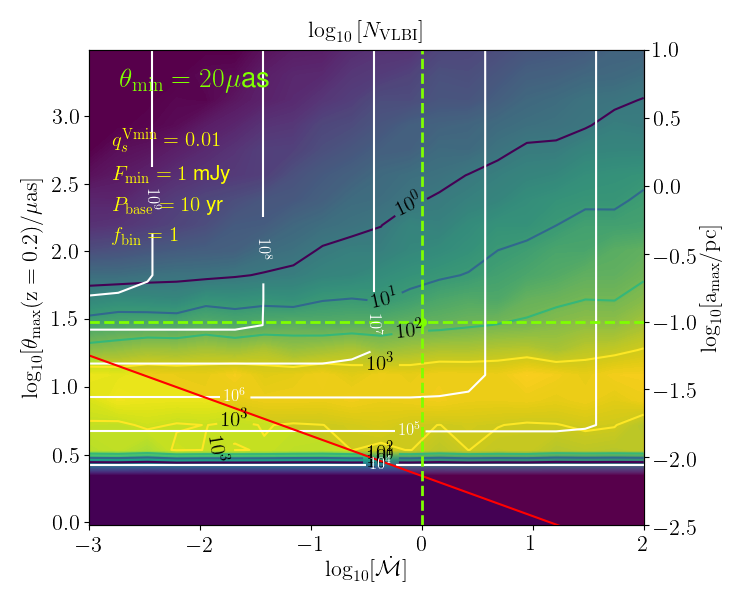} \vspace{-0.4cm} \\
\end{array}$
\end{center}
\vspace{-10pt}
\caption{
Yellow-purple shaded contours represent the Log number of resolvable MBHBs
over the entire sky, as a function of the main model parameters, namely:
$a_{\max}$, setting the orbital separation at which the binary enters a gas
disk and becomes a radio-loud AGN; $q^{\Vmin}_s$, the minimum binary symmetric
mass ratio of the resolvable population (left column); and $\mde$, which
determines the gas- driven orbital decay rate (right column).
The labeled white lines are contours of constant residence
time, in years for a $10^9 \Msun$ MBHB with $q^{\Vmin}_s=0.1$ (left column) and $\mde=1$
(right column). From top to bottom, we consider minimum instrument
resolutions of $\theta_{\min} = 1$, $10$, and $20\mu$as. The dashed-green
lines mark the chosen fiducial parameter values. In all panels the red-shaded
region indicates the possible (binary mass dependent) locations of the outer
edge of a Toomre-stable, steady-state accretion disk, which may be important
for determining the value of $a_{\max}$.
}
\label{Fig:NEHT_amx_eps}
\end{figure*}

\item \textbf{Maximum period and mass dependence:} Figure \ref{Fig:LLAGN_PvM}
illustrates the demographics of resolvable MBHBs by plotting contours of
$N_{\EHT}$ as a function of an imposed maximum observed binary orbital period
and a maximum binary mass. From top panel to bottom panel, the minimum angular
resolution is varied from $\theta_{\min}=1$ to $10$ to $20\mu$as. The white
lines show contours of constant binary separation, and the cyan lines are
contours of constant gravitational wave strain at a representative redshift of
$z=0.2$. The most massive and lowest period resolvable binaries are
approaching those that could be resolvable as individual GW sources. We
comment on this possibility in \S \ref{S:GWsources}. As expected, and pointed
out in the discussion surrounding Figure \ref{Fig:VisInt}, Figure
\ref{Fig:LLAGN_PvM} illustrates that a smaller minimum spatial resolution
implies a population of resolvable binaries with periods and masses that
extend to lower values.

The smallest binary masses at a maximum orbital period of $\sim 10$ years are
$3 \times 10^6 \Msun$ for $\theta_{\min} = 1 \mu$as, $\sim 3 \times 10^7
\Msun$ for $\theta_{\min} = 10 \mu$as, and $\sim10^8 \Msun$ for $\theta_{\min}
= 20 \mu$as. The minimum observed binary periods for the largest,
$10^{10}\Msun$, MBHBs are $\sim 1.5$ years for $\theta_{\min} = 1 \mu$as,
$\sim 3$ years for $\theta_{\min} = 10 \mu$as, and $\sim 5$ years for
$\theta_{\min} = 20 \mu$as.

\end{itemize}

\begin{figure}
\begin{center}$
\begin{array}{c}
\includegraphics[scale=0.43]{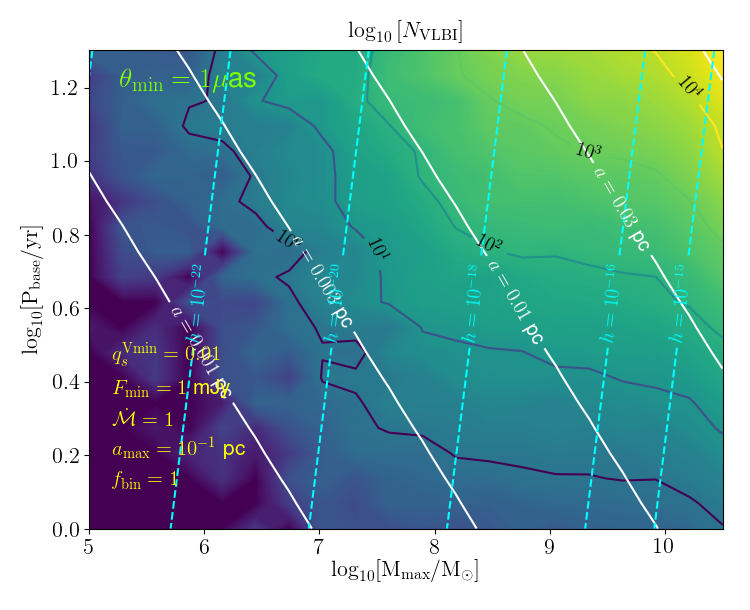} \\
\includegraphics[scale=0.43]{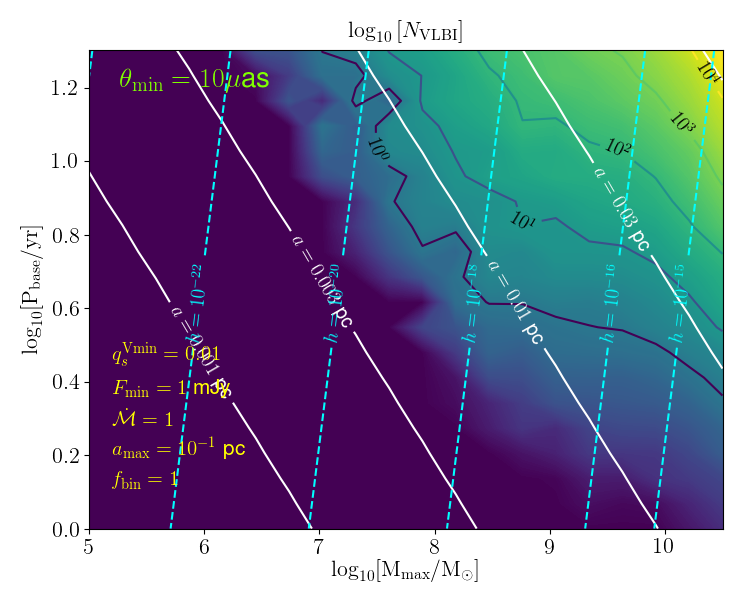} \\
\includegraphics[scale=0.43]{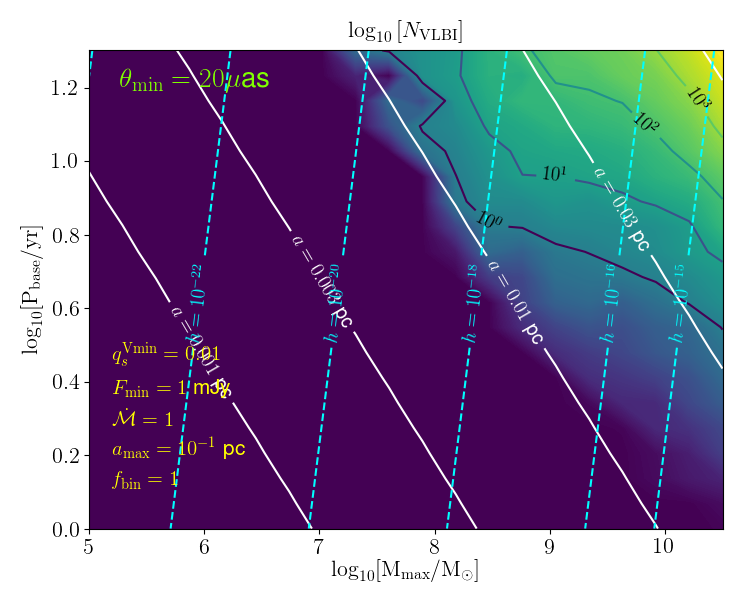} 
\end{array}$
\end{center}
\vspace{-10pt}
\caption{
Log contours of the number of resolvable MBHBs over the entire sky, as a
function of the maximum temporal baseline (maximum observed binary period) and
the maximum binary mass. From top to bottom we vary the minimum angular
resolution $\theta_{\min}$ from 1, to 10, to 20 $\mu$as. White lines are
contours of constant binary separation. Cyan lines are contours of constant
(dimensionless) gravitational-wave strain.
}
\label{Fig:LLAGN_PvM}
\end{figure}

\section{Discussion}
\label{S:Discussion}

\subsection{MBHB population}

VLBI imaging of a MBHB over the course of an orbit could provide the first
definitive proof of a MBHB at sub-pc separations. It will also probe a regime
of MBHB evolution inaccessible to GW observations. Sub-pc separations
represent the poorly understood stage in MBHB evolution where GW radiation
plus environmental effects are competing for dominance in binary orbital
decay, but where GW emission is not yet loud enough to identify the system
as an individually resolved GW source with observatories
such as the PTAs or LISA \citep{LISA:2017}. Hence, not only is this regime one
of the most important to understand for deciphering, \textit{e.g.}, the final-
parsec problem, but it is also directly accessible only in the EM sector.

VLBI imaging could attain the definitive detection of multiple MBHBs,
but importantly, work as a verification tool for more indirect methods of MBHB
identification. Linking indirect MBHB signatures, such as quasar periodicity,
variable broad lines, or jet morphology with a secure detection of a MBHB via
imaging would aid in the confident use of these indirect, but more easily
employed identification methods. Identification of at least a few MBHB systems
with VLBI imaging could even aid in identifying yet undiscovered signatures
of MBHBs in galactic nuclei.

If a population of MBHBs can be identified, the work presented in \S
\ref{S:Pdeps} can be inverted to infer the binary residence times given the
observed number of MBHBs at given separations and redshifts. In essence, a
plot similar to Figure \ref{Fig:N_vs_thmn} could be made from
observations. From this, models for the residence times can be ruled
out, and the parameters within these models can be constrained. This would
provide a powerful approach to understanding the mechanisms that drive MBHBs
through the final pc to centi-pc separations of their existence.

\begin{figure}
\begin{center}$
\includegraphics[scale=0.34]{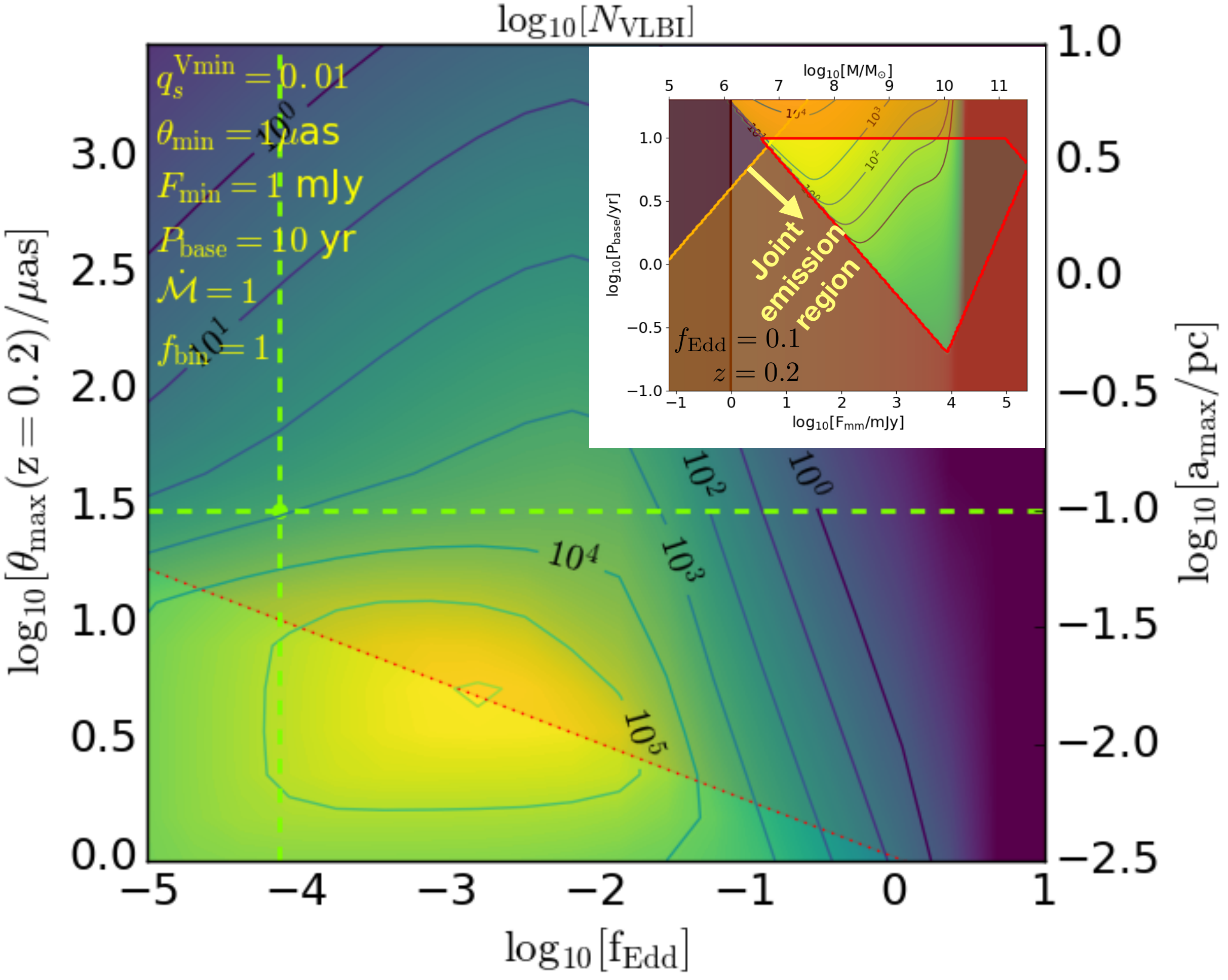} \vspace{0.25cm}
$
\end{center}
\vspace{-20pt}
\caption{
Same as the panels in Figure \ref{Fig:NEHT_amx_eps} except that the horizontal
axis varies different values of a delta function distribution of the Eddington
ratio $f_{\Edd}$. The fall off in $N_{\EHT}$ for larger $f_{\Edd}$ is due to
the growing size of the  mm-wavelength emission region. Only for low-Eddington
fractions is the mm-emission region smaller than the characteristic binary
separations considered here. The dashed-green lines mark the fiducial value of
$a_{\max}$ and the expectation value of $f_{\Edd}$ from the Eddington fraction
distribution (Eq. \ref{Eq:fEddDist}). The top-right inset is identical to the
left-middle panel of Figure \ref{Fig:VisInt} but for $f_{\rm{Edd}}=0.1$ and
without enforcing emission region size constraints.
}
\label{Fig:NEHT_amx_fEdd}
\end{figure}

The MBHB population that can be accessed through the methods proposed here
will be preferentially low Eddington ratio systems. Figure
\ref{Fig:NEHT_amx_fEdd} shows the dependence of $N_{\EHT}$ on the Eddington
ratio in the case that $f_{\Edd}$ takes on a single value. The most striking
feature of Figure \ref{Fig:NEHT_amx_fEdd} is the steep fall off of $N_{\EHT}$
for $f_{\Edd} \gtrsim 0.1$. As alluded to in the discussion surrounding Figure
\ref{Fig:VisInt}, this fall off is due solely to our criteria that the 
mm-wavelength emission region be smaller than the binary separation.

The inset of Figure \ref{Fig:NEHT_amx_fEdd} shows how the size of the  mm-
wavelength emission region grows to envelope the binary for large Eddington
fractions. This inset is identical to the left-middle panel of Figure
\ref{Fig:VisInt} except that we have plotted the integrand of Eq.
(\ref{Eq:NEHT}) (labeled contours) without imposing the emission region size
cut. We shade in yellow though where the cut would be enforced. These brighter
MBHB systems will likely not be resolved as two point sources, however we
should not preclude them as interesting targets for VLBI imaging. These
circumbinary mm-emission systems could exhibit interesting morphology or time
dependent behavior due to the binary orbital motion. It would be interesting
if this $\mu$as morphology could be linked to larger scale jet morphology.

The inset of Figure \ref{Fig:NEHT_amx_fEdd} also shows that any near-Eddington
sources that can be resolved as two individual sources will preferentially be
low mass long period binaries. This means that the resolvable MBHB population
would have a mass and period correlation with Eddington ratio.

\begin{figure}
\begin{center}$
\hspace{-15pt}
\includegraphics[scale=0.44]{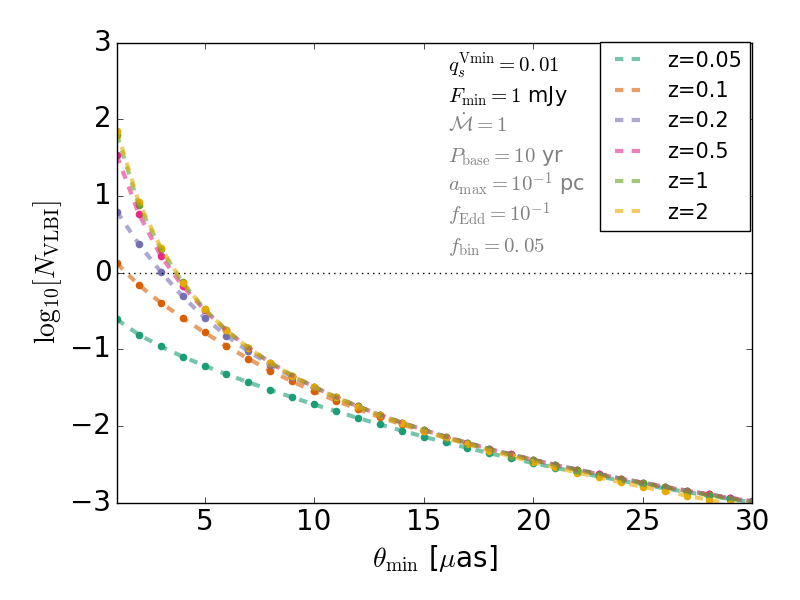} 
$
\end{center}
\vspace{-20pt}
\caption{
The same as the top left panel of Figure \ref{Fig:N_vs_thmn}, except for a
delta function distribution of $f_{\Edd}=0.1$ and not removing MBHBs with
emission regions surrounding the binary.
}
\label{Fig:NEHT_fEdd0p1}
\end{figure}

Figure \ref{Fig:NEHT_fEdd0p1} shows the distribution of such circumbinary  mm-
emission systems assuming a single Eddington ratio of $f_{\Edd}=0.1$ and not
excluding systems with mm-emission regions larger than their separation. We
see that at minimum angular resolutions approaching $1 \mu$as,
$f_{**}f_{\bin}2 \times 10^{3}$ of such systems could exist out to redshifts
of $z\sim0.5$. Future work should clarify what is expected from the mm-
emission of these systems. It is intriguing to note that the MBHB candidates
PG 1302 \citep{Graham+2015a} and OJ 287 \citep{Valtonen+2008} fall in this
latter circumbinary mm-emission region category, with angular separations of
$4\mu$as and $12\mu$as, using estimated total binary masses
of $10^{9.4}\Msun$ and $10^{10.3}\Msun$ for PG 1302 and OJ287, respectively.

\subsection{Binary Mass Determination}
\label{S:MassMeas}

Beyond probing the MBHB population, VLBI imaging of a MBHB orbit would allow a
precise measurement of the binary mass. Consider resolving the orbital
separation of a MBHB with each component active. If the binary is on a
circular orbit, its separation $a$ can be measured from the maximum resolved
angular separation $\theta_{a}$ of the two orbiting sources and the redshift
of the host galaxy $a = \theta_{a} D_A (z)$, where $D_A(z)$ is the angular
diameter distance of the source given the redshift $z$. By
tracking the binary for a large enough fraction of an orbit needed to fit an
orbital solution (this need not be an entire orbit), the binary orbital
period $P$ can be used to measure the total mass of the binary,
\begin{equation}
M =  \frac{1}{G}\left(\frac{2 \pi}{P(1+z)}\right)^2 (\theta_{a} D_A(z))^3,
\label{Eq:MandH0}
\end{equation}
where the observed period is $P(1+z)$.

Assuming Gaussian errors on the measurements of $P$, $\theta_{a}$, and
comparatively negligible errors on $z$ and the Hubble constant $H_0$ (which
factors into $D_A(z)$), the uncertainty in this mass measurement is
\begin{align}
\frac{\delta M}{M} &\approx 
 \left[\left(2 \frac{\delta P}{P}\right)^2 + \left(3 \frac{\delta \theta_{a}}{\theta_{a}} \right)^2\right]^{1/2}.
\end{align}

We estimate the error in the measurement of $\theta_{a}$ by the minimum
instrument angular resolution. When we are limited by a minimum angular
resolution of $\sim 10 \mu$as, the calculations of the previous section showed
that most resolvable binaries are at this limiting angular separation and so
$\delta \theta_{a}/\theta_{a} \sim 1$. If however, the minimum
resolution can reach $\sim 1 \mu$as, we find that $\delta
\theta_{a}/\theta_{a} \sim 0.1$.

From VLBI astrometry alone, the error in $P$ is set by the cadence of
observations $\Delta T_{\obs}$ and also the precision at which the centroid of
emission from each binary component can be determined, 
\begin{equation}
\delta P \approx \left[ \Delta T^2_{\obs} +  \left(P \frac{\theta_{\min}}{\theta_{a}}\right)^2 \right]^{1/2},
\end{equation}
which in the worst case scenario can be of order $P$. However, as we will
discuss below, identification of VLBI resolvable candidates can be carried out
through searches for quasar periodicity caused either by the relativistic
Doppler boost or variable circumbinary accretion. In these cases, the binary
period can be identified to within a few percent
\citep[\textit{e.g.}][]{Graham+2015b} \footnote{The strongest observed periodicity could be higher or lower than the orbital period \citep{PG1302-Maria, PG1302MNRAS:2015a}, but in each scenario, the true orbital period could still be discernible with a long enough observation.}. Taking into account the above best and
worse case scenarios for $\delta
\theta_{a}/\theta_{a}$, and a best case $\delta P/P \sim 0.05$, we estimate the
precision in the MBHB mass measurement to fall between
\begin{align}
 0.3 \lesssim \left.\frac{\delta M}{M}\right|_{\rm{VLBI}} \lesssim  4 .
\end{align}

The present-day state of the art technique for measuring MBH masses,
reverberation mapping, can typically measure central MBH masses to within
$0.5$ dex, or $\delta M/M|_{\rm{RM}} \sim 3$ \citep[\textit{e.g.}][and
references therein]{Shue:2013:RM}, hence the mass measurement put forth here
would rival the precision of those found through reverberation mapping
techniques, and in the best case scenario provide the most precise MBH masses
to date.Furthermore, the mass measurement proposed here is
much cleaner in that it only requires Newtonian orbit fitting, and does not rely upon
the unknown geometric factors related to AGN broad-line regions, which
contribute much of the uncertainty in the reverberation mapping analysis. The
mass measurements found via VLBI orbit-tracking could probe MBHBs out to
redshifts of $z\sim0.5$, in a mass range $10^6 \rightarrow 10^{10} \Msun$.

\subsection{Determination of the Hubble Constant}
\label{S:Hubble}

If instead there is a measurement of the central binary mass, VLBI orbit-tracking of
a MBHB allows a novel measurement of the Hubble constant. This can be achieved
by solving Eq. (\ref{Eq:MandH0}) for the Hubble constant, which determines
$D_A(z)$. As for the mass measurement of the previous section, one must again
know the redshift of the MBHB host galaxy. Using the same best case estimates
for $\delta \theta_a/\theta_a \sim 0.1$ and $\delta P /P \sim 0.05$ as in the
previous sub-section, and choosing an optimistic $\delta M /M \sim 0.5$, the
uncertainty in the Hubble constant measured from VLBI orbit-tracking is approximately,
\begin{equation}
\frac{\delta H_0}{H_0} \gtrsim 0.2 \left[ \left(\frac{\delta \theta_a / \theta_a}{0.1}\right)^2 +  \frac{4}{9}\left(\frac{\delta P/ P}{0.05}\right)^2 +  \frac{1}{9}\left(\frac{\delta M / M}{0.5}\right)^2 \right]^{1/2},
\label{Eq:H0unc1}
\end{equation}
down to $20\%$ relative error.  If the relative error in $\theta_a$ and $M$
could be decreased to $5\%$, the measurement uncertainty in $H_0$ would drop
to $6\%$.

If the binary generates periodically variable continuum emission due to the
relativistic Doppler boost \citep{PG1302Nature:2015b}, then simultaneous VLBI
monitoring of MBHB astrometry and Doppler-boosted fluxes can determine the
Hubble constant even without a binary mass measurement.

For the range of binary periods and masses for which we predict more than one
resolvable MBHB (\textit{e.g.}, the left panel of Figure \ref{Fig:LLAGN_PvM}),
the orbital velocity can range from $\sim 0.01c$, for $M\sim 10^7 \Msun$ and
$P\sim10$ yrs, up to $\sim 0.19c$ for $M\sim 10^{10} \Msun$ and $P\sim1.5$
yrs, meaning that significant Doppler modulation, at the tens of percent
level, is possible.

Assume again that we have observations of a resolved MBHB with both MBHs
active. Then mm-VLBI can observe the orbital motion projected on to the sky.
The observed angular velocity of the $i^{th}$ binary component is,
\begin{align}
\dot{\theta}_i = (1+z)\frac{v_{i}(q)}{D_A(z)} \sqrt{\cos^2{\left[\Omega t\right]} \sin^2{I} + \sin^2{\left[\Omega t\right]}},
\end{align}
where $I$ is the inclination of the binary orbital plane to the line of sight, $\Omega=2 \pi/P$ is the observed angular orbital frequency of the binary, and $v_{i}$ is the rest frame orbital velocity of the $i^{th}$ MBH. We assume a binary with mass ratio $q$ on a circular orbit,
\begin{equation}
v_p = \frac{q}{1+q} \left( \frac{GM \Omega} {1+z} \right)^{1/3} = q v_s, 
\label{Eq:vkep}
\end{equation}
where $s$ denotes secondary and $p$ denotes primary, and 
\begin{align}
D_A(z) = \frac{c}{H_0 (1+z)} \int^z_0{ \frac{dz'}{ \sqrt{ \Omega_M (1+z')^3  + \Omega_{\Lambda} }}}
\end{align}
is the angular diameter distance of the source at redshift $z$ for Hubble
constant $H_0$, and matter and dark energy density parameters $\Omega_M$ and
$\Omega_{\Lambda}$.

If the binary is on a circular orbit, the inclination of
the binary can be discerned from the projected shape of the orbit alone. The
inclination can also be recovered for an eccentric orbit. However, the change
in proper motion along the path of the eccentric orbit must also be included
to break the degeneracy between, for example, a face on elliptical orbit and a
tilted circular orbit. Assuming a circular orbit for simplicity of
demonstration,
\begin{equation}
I = \tan^{-1}\left( \frac{\theta^b}{\theta^a} \right),
\label{Eq:Inc}
\end{equation}
where $\theta^a$ and $\theta^b$ are the semi-major and semi-minor axes of the
ellipse that the binary orbital motion traces. In the more general case,
tracking the proper motion of the binary components will allow measurement of
the orbital eccentricity.

Similarly, the mass ratio could be measured if the binary orbit is resolved on
the sky. For example, a binary on a circular orbit with semi-major axis $a$,
has a secondary that orbits at a distance $r_s=a/(1+q)$ from the binary center
of mass while the primary orbits at a distance $r_p=aq/(1+q)$. The measurement
of the ratio of these angular distances yields the binary mass ratio.

The orbital velocity can be measured independently from monitoring the
periodically changing flux caused by the relativistic Doppler boost. Generally,
the emission from both MBHs must be taken into account,
\begin{widetext}
\begin{align}
\frac{ F^{\obs}_{\nu} }{ F^0_{\nu} }(t) &= f_s \left[ \gamma_s \left(1 - \frac{v_{s||}}{c} \right) \right]^{\alpha - 3}  + (1-f_s) \left[ \gamma_p \left(1 + q\frac{v_{s||}}{c} \right) \right]^{\alpha - 3} \nonumber \\
v_{s||} &= v_s \cos{\left[\Omega t\right]} \cos{I};  \quad \gamma_i = \left[1 - \left(\frac{v_i}{c}\right)^2 \right]^{-1/2},
\end{align}
\end{widetext}
where $||$ denotes the velocity component along the line of sight and we have
used that $v_{p||}= -qv_{s||}$. This introduces a new quantity $f_s = \left<
F_s/(F_s + F_p) \right>$, the fraction of time averaged flux detected from the
secondary compared to the total. From measurements of $\alpha$, $I$, $q$, and
the time dependent flux, we solve the above equation for the orbital velocity
of the secondary and primary MBH.

Hence, the two measurements of the Hubble constant are given by
\begin{align}
(H_0)_{i} =  \frac{c \dot{\theta}_i  \int^z_0{ \frac{dz'}{ \sqrt{ \Omega_M (1+z')^3  + \Omega_{\Lambda} }}} }{ v_{i}  \sqrt{ \cos^2{\left[\Omega t\right]} \sin^2{I} + \sin^2{\left[\Omega t\right]} } }.
\end{align}
which must be averaged over some fraction of the binary orbit in order to
measure $\dot{\theta}_i$. This measurement of $H_0$ could then be used in
conjunction with the method using Eq. (\ref{Eq:MandH0}) to further improve the
uncertainty in the VLBI-measured Hubble constant.

Note that this method for measuring the Hubble constant with proper motions
and orbital velocities can also be carried out without detection of the
Doppler boost, but with a measurement of the binary mass and mass ratio. In
this case, the orbital velocity of each MBH yields via Eq. (\ref{Eq:vkep}).

As an example for the precision at which the Hubble constant can be measured in
the Doppler-boost method, we envision the simplest case. We consider the case where
only the Doppler boost from the secondary is important and where $\gamma_s \sim
1$.\footnote{This applies, for example, in the case that VLBI can isolate emission from
one of the binary components, or when the mass ratio is disparate and only the
secondary contributes strongly to the Doppler boost.} Then the secondary
orbital velocity can be written as,
\begin{eqnarray}
\frac{v_s}{c}\cos{I} = \left[\Delta_{+} + 1 \right]^{\frac{1}{\alpha-3}} - 1
\\ \nonumber
\Delta_{+} \equiv \frac{ F^{\obs}_{\nu}(0)}{F^0_{\nu}} - \frac{ F^{\obs}_{\nu}(P/4)}{F^0_{\nu}},
\end{eqnarray}
where the $\Delta_{+}$ is the modulation amplitude between peak and
average flux (between $t=0$ and $t = P/4$). Assuming that uncertainty in the
binary inclination dominates, we can write $\delta v/v \approx I \tan{I}
(\delta I/I)$. Assuming a small inclination angle to the line of sight (needed
for the Doppler boost to be significant), we approximate $I \sim
\theta_b/\theta_a$ (from Eq. \ref{Eq:Inc}), and $\delta I /I \approx
\sqrt{2} (\delta \theta_a/\theta_a)$. Taking optimistic values of $I \leq
0.5$ rad, $\delta \dot{\theta} / \dot{\theta} = 0.1$, $\delta \theta_a / \theta_a
= 0.1$, and $\delta \Omega /\Omega = 0.05$,
\begin{equation}
\frac{\delta H_0}{H_0} \gtrsim 0.1 \qquad \mbox{(Doppler} \ \mbox{case)},
\label{Eq:H0unc1}
\end{equation}
which is valid for $I\leq0.5$, where this estimate does not vary greatly over
the course of the orbit. If these measurements can be made
out to redshifts $z>0.1$, as Figure \ref{Fig:N_vs_thmn} suggests is possible,
then they could provide an independent measure of the Hubble constant that
could aid in resolving discrepancies in other independent measurements such as
the current mismatch between $H_0$ measured by Planck via the cosmic microwave
background \citep{PlanckH0:2016}, and the value measured by the Hubble space
telescope via the Cepheid variables \citep{Riess+2016}.

\subsection{Observational Strategy}

As mm-VLBI is not suited for all-sky surveys, we require pilot
observations that can identify MBHB mm-bright candidates for VLBI follow up.
We have shown that the majority of resolvable MBHBs would be at the heart of
mm-bright LLAGN. This means that there will be only a dim optical/UV component
to these sources, (LLAGN do not exhibit a big-blue bump in their spectra).
LLAGN spectra, however, are brighter in the near-IR \citep[\textit{e.g
.}][]{Fern-Onti:LLSED+2012} and hence accessible via upcoming time-domain
surveys such as the Large Synoptic Survey Telescope \citep[LSST;][]{LSST}.
Hence we propose the following general strategy to find VLBI MBHB candidates:
\begin{enumerate}
	\item 
	Search for periodicity with all-sky time domain surveys in the near-IR,
	\textit{e.g.}, LSST. Such a survey will identify a list of MBHB candidates
	from which we can select those with the required binary separations given
	a mass estimator and the observed variability period \citep[see
	\textit{e.g.},][]{Graham+2015b, Charisi+2016}. We note that the lifetime
	of these surveys will set the size of the maximum binary period
	$P_{\rm{base}}$. Currently, surveys such as the Catalina Real Time
	Transient Survey have a 10 year temporal baseline \citep{CRTS4:Djorgovski:2011}.
	\item 

	Determine whether each periodic-light-curve candidate is radio loud (and
	hence bright in mm-wavelengths), either via archival searches or follow up
	with single radio observations.
	\item 
	Observe candidates that pass stages 1 and 2 with mm-VLBI
	and determine if they consist of two compact sources, or one compact source
	and a nearby phase calibrator. If they do, monitor these sources for orbital
	motion. Even for the longest period binaries considered here ($10$ years),
	two or three observations spaced by a year would be sufficient to test for
	orbital motion.
\end{enumerate}
We point out that an advantage to searching for sub-pc separation, dual-source
binaries, over searching for wide, pc-kpc dual AGN, is that these smaller
separation binaries could be identified for follow-up via the imprint of their
orbital period.

\subsection{Gravitational wave single-source detection}
\label{S:GWsources}

As shown in the cyan contours of Figure \ref{Fig:LLAGN_PvM}, the nearest MBHBs
at the high mass and low period end of the resolvable population could be
detectable as single GW sources\footnote{See \cite{SchutzMa:2016} for
constraints on the mass-ratios of a nearby, most-massive
population of putative MBHBs.}. In this case, the EM discovery of a MBHB
could aid the PTAs in digging out a weak signal of the binary. Such
identification could also allow a precision binary mass measurement that could
be used to increase the precision of the Hubble constant measurement of \S
\ref{S:Hubble}, or to corroborate the mass measurement of \S \ref{S:MassMeas}.
A simultaneous GW detection could also contribute a second, independent
measurement of the Hubble constant through the standard sirens approach
\citep[see][and references therein]{Schutz:1986, StandSirenGW170818:2017}.

Finally, VLBI orbit-tracking plus GW detection of a MBHB could be used to
measure the relative speed of light and GWs. This can be achieved by tracking
the binary orbital phase through both the EM and GW messengers. The detection
of the relativistic Doppler boost would further enhance the capability to make
such a measurement \citep[see][for a similar scenario]{Haiman:2017}.

\section{Conclusions and Future Work}
\label{S:Conclusions}

By constructing simple models for gas and gravitational-wave-driven binary
orbital decay and a mm-wavelength AGN luminosity function, we estimate the
number of MBHBs, over the entire sky, that reside at separations directly
resolvable by mm-wavelength VLBI, and with orbital periods of less than 10
years.  We show that, $1-10\mu$as resolution, sub-mm VLBI with flux
sensitivity better than 1 mJy can resolve the orbital separation of tens of
MBHBs at fiducial model parameters, and up to $\sim 10^4$ at parameter values
tuned to maximize this number. These MBHBs are found out to redshift
$z\approx0.5$ and have total masses above $\sim 10^7 \Msun$.

For a minimum flux sensitivity of $10$mJy, closer to current capabilities,
there are of order a few resolvable MBHBs out to redshift $z\lesssim0.2$. For
an enhanced sub-mm flux sensitivity of $0.1$mJy, our fiducial MBHB-decay model
predicts that a few $\times 10^2$ MBHBs will be resolvable at $1\mu$as
resolution out to $z\lesssim1.0$, while of order one MBHB will be resolvable
with $25 \mu$as resolution. Further extending the maximum temporal baseline to
include 20 year binary orbital periods increases the all-sky number of
resolvable MBHBs to a few thousand at the best case $1\mu$as resolution, and a
few tens even at a minimum angular resolution of $30\mu$as.

We determine that resolvable MBHBs, for which each component is tracked by a
mm-emission region smaller than the binary orbit, preferentially reside in
low-luminosity AGN. These objects could be identified through periodicity
signatures in near-IR/optical time domain surveys and followed up with
radio observations as a precursor for VLBI orbit-tracking.

Resolvable MBHBs will be emitting gravitational radiation in the PTA band
($\sim$nHz). If they are within a redshift of $0.1$, the closest, most massive
MBHBs could be detectable as individual binary sources. We show that the total
binary fraction of these low-luminosity AGN is constrained to by $\lesssim0.05$
by PTA limits on the gravitational wave background.

Beyond providing definitive existence of sub-pc MBHBs, VLBI tracking of a
complete binary orbit would allow a measurement of the binary mass
(to within $\sim30 \%$), or a novel, $\mathcal{O}(10\%)$, measurement of the
Hubble constant, if the binary mass is known.

Future work should consider improved models for the orbital decay of MBHBs and
their mm-bright lifetimes. This is important for understanding the range in
predictions for the number of resolvable MBHBs and also for determining how
well such models can be ruled out by future population estimates.
Additionally, future work should understand the population of periodic-light-
curve  candidates that could be identified with current and future time domain
surveys, specifically, the subset of these that are VLBI imaging candidates.
Further analysis of the existing MBHB candidates in the literature may already
provide interesting targets for VLBI orbit tracking. For example, PG 1302-102,
OJ 287, and 3C 273 are each sub-mm bright AGN within redshift $0.3$ that have
been reported as MBHB candidates with observed periods of $5$, $12$, and $16$
years \citep{Graham+2015a, Valtonen+2008, AbrRom:1999, Romero:2000}
respectively. From total binary mass estimates of $10^{9.4}\Msun$,
$10^{10.3}\Msun$, and $10^{9}\Msun$, this corresponds to putative angular
separations of the binary orbits of $4$, $12$, and $10\mu$as.

\acknowledgements 
The authors thank the referee, Zolt\'an Haiman, for a very insightful and
thorough review which greatly enhanced the clarity of this work. The authors also 
thank Eliot Quataert, Michael Johnson, Lindy Blackburn, Sheperd Doeleman,
Michael Eracleous, Atish Kamble, Bence Kocsis, Anna Pancoast,
and Mark Reid for useful discussions. Financial support was provided from NASA
through Einstein Postdoctoral Fellowship award number PF6-170151 (DJD) and
through the Black Hole Initiative which is funded by a grant from the John
Templeton Foundation.

\appendix

\section{Millimeter-wavelength AGN luminosity function}
\label{S:mmALF}

We use a redshift dependent luminosity function of radio-loud AGN
that takes into account both density evolution and luminosity evolution
\citep{YuanRLFII+2017},
\begin{widetext}
\begin{eqnarray}
\frac{d^2N}{dL dV} &=& e_1(z) \frac{\phi}{ L_{408}/e_2(z) \ln{10} } \left(\frac{L_{408}/e_2(z)}{L_*}\right)^{-\beta} \exp{\left[ -\frac{L_{408}/e_2(z)}{L_*}^{\gamma}\right]} \nonumber \\ \nonumber \\
e_1(z) &=& \frac{(1+z_c)^{-p1} + (1+z_c)^{-p2}}{(\frac{1+z}{1+z_c})^{p1} + (\frac{1+z}{1+z_c})^{p2} },
\qquad e_2(z) = 10^{k_1z + k_2 z^2}, \nonumber \\
L_* &=& 10^{24.79} \rm{W Hz}^{-1}, \qquad \phi = 10^{-4.72} \rm{Mpc}^{-3}, \qquad p1 = -1.29, \qquad p2 =6.80, \nonumber \\
\beta &=& 0.45, \qquad \gamma = 0.31, \qquad k_1, = 1.44 \qquad k_2 = -0.16, z_c = 0.78,
\end{eqnarray}
\end{widetext}
where $L_{408}$ is the specific luminosity at $408$ MHz. This is scaled to a
mm-wavelenght luminosity function as dscribed in the main text. We note for
ease of reproducibility, that in Table 1 of \citep{YuanRLFII+2017} the values
of $k_1$ and $k_2$ are erroneously swapped while the values of $p_1$ and $p_2$
are erroneously stated as their negatives (needed to reproduce Figures 3
and 4 in \cite{YuanRLFII+2017}).

\section{The size of the mm-wavelength emission region in LLAGN} 
\label{S:mmSize}

We require that the size of the mm-wavelengths emission region be smaller
than the binary separation. We estimate the size of the mm-emission region
with the jet model of \citep{BK79}.

The radio and mm-emission from MBHs is likely from synchrotron
emission generated by shocks in a jet. At low frequencies, synchrotron
radiation is optically thick to self absorption and its spectrum rises as
$F^{\rm{thick}}_{\nu} \propto \nu^{5/2}$ until the radiation becomes optically
thin at frequency $\nu_{ssa}$. For higher frequencies the optically thin
spectrum falls off as $F^{\rm{thin}}_{\nu} \propto  \nu^{-(p-1)/2}$ where the
electrons are assumed to be distributed as $N_e = K_e \gamma^{-p}_e$. At even
higher frequencies, however, the synchrotron losses are great enough to cool
the radiation within a jet expansion time. Then above some $\nu_{\rm{loss}}$, the
synchrotron flux drops more steeply as $F^{\rm{loss}}_{\nu} \propto
\nu^{-(p-1)/2-0.5}$.

Both limiting frequencies, and hence the shape of the synchrotron spectrum,
depends on radial position within the jet. The self absorption frequency
$\nu_{ssa}$, for which only higher frequency synchrotron photons can escape
becomes larger at larger radii in the jet. The maximum frequency for which
synchrotron losses are negligible, $\nu_{loss}$, decreases with radius in the
jet. Hence, there is a minimum jet radius $r_{\min}$ below which no bright,
optically thin synchrotron spectrum exists. This minimum radius is given by
equating $\nu_{ssa}(r)=\nu_{\rm{loss}}(r)$.

Then for the mm-emission regions to be smaller than the binary separation, we
must have, $\nu_{ssa}(a)<\nu_{mm}<\nu_{\rm{loss}}(a)$.
In practice we evaluate this inequality for a given binary separation $a$ and
set the resolvable MBHB probability $\mathcal{F}$ equal to zero if it is not
satisfied.

The self absorption frequency is given by \cite[\textit{Eq. 28 of}][]{BK79},
\begin{widetext}
\begin{equation}
\nu_{ssa} = \frac{300}{1+z} k^{1/3}_e  \left[\Delta\left(1 + \frac{2}{3} k_e \Lambda \right) \right]^{-2/3} \gamma^{-4/3}_j \beta^{-2/3}_j D^{2/3}_j (\sin{\theta})^{-1/3} \phi^{-1}_{\rm{ob}} L^{2/3}_{44} \left( \frac{r_{\rm{ob}}}{0.01 \rm{pc}} \right)^{-1} \ \mbox{GHz},
\end{equation}
where $k_e\lesssim1$, $\Delta$ is the logarithm of the ratio of maximum and
minimum size scales in the jet, $\Lambda$ is the logarithm of the ratio of
maximum and minimum electron Lorentz factors in the jet, $\gamma_j = \left(1 -
\beta^2_j\right)^{-1/2}$ is the Lorentz factor of the jet, $D_j$ is the jet
Doppler factor, $\theta$ is the viewing angle  of the jet, $\phi_{\rm{ob}}=
\phi/\sin{\theta}$ is the observed opening angle of the jet, $L_{44}$ is the
bolometric luminosity of the jet in units of $10^{44}$ erg s$^{-1}$, and
$r_{\rm{ob}}=r\sin{\theta}$ is the observed distance from jet launching point
to the emission region. This suggests that sources with $L \sim 10^{44}$ erg
s$^{-1}$ are optically thin at mm-wavelengths down to size scales of $r \sim
0.01$pc. Lower luminosity sources are visible down to even smaller scales.

The maximum frequency for which synchrotron losses are negligible,
\begin{equation}
\nu_{loss} = \frac{0.07}{1+z} \gamma^{2}_j \beta^2_j \frac{D_j}{\sin{\theta} B^3_1} r_{\rm{ob}},
\end{equation}
where $B_1$ is the magnetic field strength at $r=1$pc,
\begin{equation}
B_1 = 2  \left[\Delta\left(1 + \frac{2}{3} k_e \Lambda \right) \right]^{-1/2} \phi (\beta_j c)^{-1/2} \gamma_j \left( \frac{r}{ 1 \rm{pc}} \right)^{-1} \left( \frac{L}{ 10^{44} \rm{erg} \ \rm{s}^{-1} } \right)^{1/2},
\end{equation}
\end{widetext}
is found by equating the total power in the jet to that carried away by
relativistic electrons and the magnetic field \citep[\textit{Eq. 23
of}][]{BK79}.

Throughout we use $k_e=1$, $\Delta = \Lambda = \ln(10^5)$,  $\gamma_j=10$,
$\phi=1/\gamma_j$, and $\theta = 0.1$. The choice of $\theta$ is
conservative as it yields the maximum value of $\nu_{ssa}$.

\bibliography{refs}
\end{document}